\documentclass{article}

\usepackage{biblatex}

\usepackage{amsmath}
\usepackage[nopatch]{microtype}
\usepackage{booktabs}
\usepackage{longtable}

\usepackage{pgfplots}

\usepackage{arydshln}

\usepackage{amsthm}
\usepackage{amssymb,verbatim}

\usepackage{todonotes}

\usepackage{hyperref}

\usepgfplotslibrary{ternary}

\usepackage{graphicx}

\newcommand{\Conv}{\mathop{\scalebox{2.5}{\raisebox{-0.2ex}{$\ast$}}}}

\newcommand{\triplechart}[2]{#2} 

\theoremstyle{definition}
\newtheorem{lem}{Lemma}

\newtheorem{exl}[lem]{Example}

\newcommand{\R}{\mathbb R}

\newcommand{\cE}{\mathcal E}
\newcommand{\cC}{\mathcal C}

\newcommand{\winplotwidth}{7cm}
\newcommand{\tiledchartwidth}{4cm}

\title{Ahead of the Count: An Algorithm for Probabilistic Prediction of Instant Runoff (IRV) Elections}

\author{Nicholas Kapoor and P. Christopher Staecker
\footnote{Department of Mathematics, Fairfield University, Fairfield CT, USA, \texttt{nicholas.kapoor@fairfield.edu}, \texttt{cstaecker@fairfield.edu}}}

\begin{document}

\maketitle

\begin{abstract}
How can we probabilistically predict the winner in a ranked-choice election without all ballots being counted? In this study, we introduce a novel algorithm designed to predict outcomes in Instant Runoff Voting (IRV) elections. 
The algorithm takes as input a set of discrete probability distributions describing vote totals for each candidate ranking and calculates the probability that each candidate will win the election. In fact, we calculate all possible sequences of eliminations that might occur in the IRV rounds and assign a probability to each.

The discrete probability distributions can be arbitrary and, in applications, could be measured empirically from pre-election polling data or from partial vote tallies of an in-progress election.

The algorithm is effective for elections with a small number of candidates (five or fewer), with fast execution on typical consumer computers. The run-time is short enough for our method to be used for real-time election night modeling where new predictions are made continuously as more and more vote information becomes available. We demonstrate the algorithm in abstract examples, and also using real data from the 2022 Alaska state elections to simulate election-night predictions and also predictions of election recounts. 
\end{abstract}





\section{Introduction} \label{introsection}
The American electorate is antsy. The build-up of an election season is palpable when polls close on election night; the need to know the winner is insatiable. The analysis of the electorate’s decisions – before they are known – begins immediately with exit polling and the first trickles of vote tallies. Due to different state laws around counting the vote, tallies are produced at varying rates, in some cases taking several days.



Regardless of the speed of counting the votes, different news agencies begin predicting the winner. In a traditional plurality voting system (sometimes called first-past-the-post, or FPTP), predicting the winner based on polling data and other models is fairly straightforward from a mathematical point of view. In this paper, we give a method for predicting the winner in elections using Hare Instant Runoff Voting (IRV), commonly called ``ranked choice voting" (RCV) in the US.

Initiatives to replace FPTP elections with IRV elections have had 27 consecutive wins on municipal ballots \autocite{otis_ranked_nodate}, and this movement is quickly becoming one of the largest voting reforms in the United States. As IRV becomes a larger part of American politics, new tools are required for analysts who want to predict election outcomes probabilistically. For FPTP elections, for example, \textit{The New York Times} uses their infamous ``needle'' that tracks in real-time how likely one candidate is to defeat the other as the votes are being counted. Our paper will present a basic mathematical framework for the design of a similar ``needle'' for IRV elections. More generally, our work can be used to make probabilistic predictions of IRV elections long before Election Day based on polling or other data, or it can even be applied after elections to make predictions of recount results. 

The 2019 New York City Mayoral and City Council Democratic primaries were one of the first major uses of IRV in the US in multiple races on the same Election Day. \textcite{a_jelvani_identifying_2022} provide an algorithm to determine the possible elimination outcomes in the 2019 New York City Democratic Mayoral primary based on incomplete tallying of the ballots. Our paper can be seen as an extension of this work to a probabilistic setting.

Another related work is \textcite{bhattacharyya_2021}, which gives methods using a sample of votes to predict the winner of an election probabilistically. That paper examines many different voting methods, not just IRV, and focuses on the question of how large the sample must be in order to make robust predictions. Similarly, our work can also make predictions of elections where only a sample of votes are known: this is the main point of view in the examples of Section \ref{unboundsec}.

In this paper, we present an algorithm which calculates the set of possible outcomes of an IRV election and assigns probabilities to each one.

The input to the algorithm is a set of discrete probability density functions describing anticipated vote totals for each possible candidate ranking. 
For example, in an election with three candidates, A, B, and C, these probability distributions might look like \footnote{For example, the third column above indicates that the ranking $BA$ will receive 200 votes with probability $45\%$, 300 votes with probability $31\%$, etc. Dots represent the probability of 0.}:
\[
\begin{tabular}{c|ccccccccc}
Votes & $f_{A}$ & $f_{B}$ & $f_{C}$ & $f_{AB}$ & $f_{AC}$ & $f_{BA}$ & $f_{BC}$ & $f_{CA}$ & $f_{CB}$ \\
\hline 
0 & 0.50 & 0.10 & 0.02 & 0.01 & 0.50 & . & . & 0.09 & 0.17 \\
100 & 0.50 & 0.30 & 0.33 & 0.17 & 0.40 & . & 0.10 & 0.17 & 0.75 \\
200 & . & 0.30 & 0.21 & 0.34 & 0.07 & 0.45 & 0.30 & 0.37 & 0.08 \\
300 & . & 0.20 & 0.20 & 0.25 & 0.03 & 0.31 & 0.27 & 0.21 & . \\
400 & . & 0.10 & 0.15 & 0.13 & . & 0.20 & 0.19 & 0.10 & . \\
500 & . & . & 0.09 & 0.10 & . & 0.04 & 0.14 & 0.06 & . \\
\end{tabular}
\]

We note that our method makes no assumptions on the shape of the individual probability distributions. There seems to be some debate in the literature on the question of which distribution functions will most accurately model voting results. The obvious candidate is the normal distribution, but some have suggested that this is unrealistic (Gelman, Katz, and Tuerlinckx 2002; Dow and Enderbsy 2004; Bovens and Biesbart 2011). We take no position on the matter, and our method can be applied to any imagined probability distribution function, either standard mathematical distributions (e.g. the normal distribution), or arbitrary distributions constructed from empirical measurements.

From the table above, our algorithm calculates probabilities for victory by each of A, B, and C. In this example, we compute winning probabilities as follows:
\[ A: 4.8\% \qquad B: 86.1\% \qquad C: 8.9\% \]
Full details of the calculation are presented later in Example \ref{sampleexample}.

In fact, our algorithm gives the probabilities associated with every possible round in the IRV calculation. In this example, we can present the probabilities of the various rounds in a ``weighted elimination tree'' as follows:
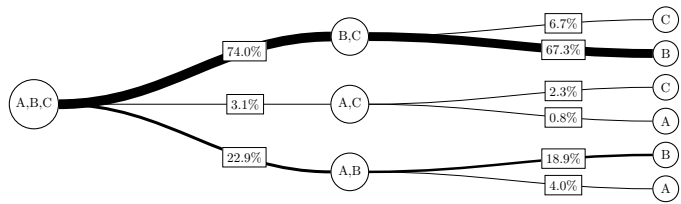
\[
\tikzset{vert/.style={draw=black,circle}}
\tikzset{weight/.style={fill=white,line width=.1,draw=black,rectangle,pos=.7}}
\begin{tikzpicture}[scale=0.6,every node/.style={scale=0.5}]
\node[vert] (ABCx) at (0,0) {A,B,C};
\node[vert] (BCxA) at (7,1.5) {B,C};
\node[vert] (CxAB) at (14,1.875) {C};
\draw[line width=0.3347833558926131] (BCxA) to[out=0,in=180] node[weight] {6.7\%} (CxAB);
\node[vert] (BxAC) at (14,1.125) {B};
\draw[line width=3.3664669099948124] (BCxA) to[out=0,in=180] node[weight] {67.3\%} (BxAC);
\draw[line width=3.701250265887426] (ABCx) to[out=0,in=180] node[weight] {74.0\%} (BCxA);
\node[vert] (ACxB) at (7,0.0) {A,C};
\node[vert] (CxBA) at (14,0.375) {C};
\draw[line width=0.11425619503727799] (ACxB) to[out=0,in=180] node[weight] {2.3\%} (CxBA);
\node[vert] (AxBC) at (14,-0.375) {A};
\draw[line width=0.03902805354479181] (ACxB) to[out=0,in=180] node[weight] {0.8\%} (AxBC);
\draw[line width=0.1532842485820698] (ABCx) to[out=0,in=180] node[weight] {3.1\%} (ACxB);
\node[vert] (ABxC) at (7,-1.5) {A,B};
\node[vert] (BxCA) at (14,-1.125) {B};
\draw[line width=0.9432595944273449] (ABxC) to[out=0,in=180] node[weight] {18.9\%} (BxCA);
\node[vert] (AxCB) at (14,-1.875) {A};
\draw[line width=0.20220589110316017] (ABxC) to[out=0,in=180] node[weight] {4.0\%} (AxCB);
\draw[line width=1.1454654855305049] (ABCx) to[out=0,in=180] node[weight] {22.9\%} (ABxC);
\end{tikzpicture}
\]
Above, each node represents a round involving the labeled candidates, and the weighted edges represent the probability of each round following from the previous. For example, the three percentages on the left are the probabilities of each of A, B, and C being eliminated in the first round. The $67.3\%$ at the upper right is the probability that B wins the election after A was eliminated first. This number, plus the $18.9\%$ below, adds up to the total probability of a win by B. For readability, the thickness of each edge is drawn in proportion to its probability.

This tree provides a probabilistic analog of similar diagrams from \textcite{a_jelvani_identifying_2022}, for example, their Figure 1 which calculates which eliminations are mathematically possible but does not indicate their probabilities.

\section{Sample applications of our algorithm}\label{alaskasection}
After adopting Ballot Measure 2 in 2020, Alaska was the first state to implement a top-four primary followed by an IRV general election. All qualifying candidates are on a single primary ballot where voters vote for their top choice only. The top four vote-getters in the primary then proceed to an IRV general election. Ballot Measure 2 passed with 50.55\% vote in 2020 \autocite{Brooks_2020} and has survived a court challenge in the Alaska Supreme Court \autocite{kohlhaas_2022}. However, a ballot measure to repeal IRV elections in Alaska may make the 2024 ballot \autocite{samuels_2024}. 

We will present some examples modeled on real data from the 2022 General Election in the state of Alaska, which used IRV for all contests. The complete voting data for these elections is publicly available online from the Alaska  Division of Elections.\footnote{\url{https://bit.ly/Alaska2022GE}}

\subsection{Winner prediction based on partial vote counts}\label{unboundsec}
\begin{exl} {\bf 2022 Alaska House District 18.}\label{house18exl}


As a fairly straightforward example, we first consider the 2022 election for Alaska House District 18, in which the ballot listed Democrat and eventual winner Cliff Groh, Democrat Lyn Franks, and Republican David Nelson. We denote these candidates respectively by G, F, and N.

In the voting data published by the state of Alaska, each vote comes tagged with a geographical identifier called the ``precinct portion,'' which tracks where the vote originated. The 2022 election data lists votes from about 500 different precinct portions. 

We will apply our algorithm to a hypothetical election-night scenario in which only a portion of the ballots have been counted. Taking the full set of around 2,{}100 votes cast in the House District 18 election, we choose 50\% of these votes and tally them precisely.

In this particular instance, after counting 50\% of the votes, our tally is as follows:

\begin{equation}\label{FGNvotes}
\begin{tabular}{r|ccccccccccc}
ranking: &  F &  N &  G &  FN &  FG &  NF &  NG &  GF &  GN & \\
\hline
votes: &  55 &  301 &  96 &  20 &  147 &  38 &  52 &  245 &  42 & \\
\end{tabular}
\end{equation}


For the remaining 50\% of the votes, we must make a probabilistic prediction of their contents. In a real-life application of our algorithm, this prediction could be modeled using pre-election polling data, ideally coupled with geographic and other information which allows more precise predictions for the particular group of votes which are still uncounted.

As a simple proof-of-concept, without access to real polling data, we simply predict that the uncounted ballots will be tallied roughly in proportion to those that have already been counted. Specifically, for some ranking $\ell$, we set the probability distribution function $f_\ell$ equal to a discretized normal distribution with mean and standard deviation in proportion to the mean and standard deviation of the already-counted ballots.

In this case, our probability distributions are as follows:

\[
\begin{tabular}{c|ccccccccc}
Votes & $f_{F}$ & $f_{N}$ & $f_{G}$ & $f_{FN}$ & $f_{FG}$ & $f_{NF}$ & $f_{NG}$ & $f_{GF}$ & $f_{GN}$ \\
\hline 
0 & . & . & . & 1.00 & . & . & . & . & . \\
75 & 0.39 & . & 0.21 & . & . & 0.49 & 0.40 & . & 0.45 \\
150 & 0.61 & . & 0.33 & . & 0.14 & 0.51 & 0.60 & . & 0.55 \\
225 & . & . & 0.29 & . & 0.20 & . & . & 0.08 & . \\
300 & . & 0.07 & 0.14 & . & 0.23 & . & . & 0.11 & . \\
375 & . & 0.09 & 0.04 & . & 0.20 & . & . & 0.13 & . \\
450 & . & 0.10 & . & . & 0.13 & . & . & 0.14 & . \\
525 & . & 0.11 & . & . & 0.07 & . & . & 0.14 & . \\
600 & . & 0.11 & . & . & 0.03 & . & . & 0.12 & . \\
675 & . & 0.11 & . & . & . & . & . & 0.10 & . \\
750 & . & 0.10 & . & . & . & . & . & 0.07 & . \\
825 & . & 0.09 & . & . & . & . & . & 0.05 & . \\
900 & . & 0.07 & . & . & . & . & . & 0.03 & . \\
975 & . & 0.05 & . & . & . & . & . & 0.02 & . \\
1050 & . & 0.04 & . & . & . & . & . & 0.01 & . \\
1125 & . & 0.03 & . & . & . & . & . & . & . \\
1200 & . & 0.02 & . & . & . & . & . & . & . \\
1275 & . & 0.01 & . & . & . & . & . & . & . \\
1350 & . & 0.01 & . & . & . & . & . & . & . \\
\end{tabular}
\]

The algorithm we will describe in Section \ref{algorithmsection} computes the probabilities of each candidate winning the election as follows:
\[
 \qquad F: 2.6\%  \qquad N: 25.3\%  \qquad G: 72.1\% 
\]
with the following weighted elimination tree:
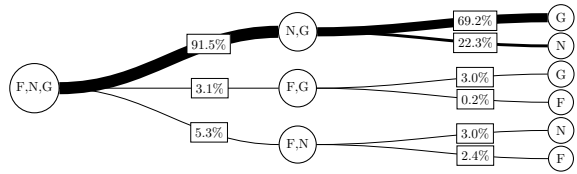
\[
\tikzset{vert/.style={draw=black,circle}}
\tikzset{weight/.style={fill=white,line width=.1,draw=black,rectangle,pos=.7}}
\begin{tikzpicture}[scale=0.5,every node/.style={scale=0.5}]
\node[vert] (505152x) at (0,0) {F,N,G};
\node[vert] (5152x50) at (7,1.5) {N,G};
\node[vert] (52x5051) at (14,1.875) {G};
\draw[line width=3.4595352624535556] (5152x50) to[out=0,in=180] node[weight] {69.2\%} (52x5051);
\node[vert] (51x5052) at (14,1.125) {N};
\draw[line width=1.1170632875108428] (5152x50) to[out=0,in=180] node[weight] {22.3\%} (51x5052);
\draw[line width=4.576598549964398] (505152x) to[out=0,in=180] node[weight] {91.5\%} (5152x50);
\node[vert] (5052x51) at (7,0.0) {F,G};
\node[vert] (52x5150) at (14,0.375) {G};
\draw[line width=0.1476696954039419] (5052x51) to[out=0,in=180] node[weight] {3.0\%} (52x5150);
\node[vert] (50x5152) at (14,-0.375) {F};
\draw[line width=0.009070610338145949] (5052x51) to[out=0,in=180] node[weight] {0.2\%} (50x5152);
\draw[line width=0.15674030574208783] (505152x) to[out=0,in=180] node[weight] {3.1\%} (5052x51);
\node[vert] (5051x52) at (7,-1.5) {F,N};
\node[vert] (51x5250) at (14,-1.125) {N};
\draw[line width=0.147974710242919] (5051x52) to[out=0,in=180] node[weight] {3.0\%} (51x5250);
\node[vert] (50x5251) at (14,-1.875) {F};
\draw[line width=0.11868643405059555] (5051x52) to[out=0,in=180] node[weight] {2.4\%} (50x5251);
\draw[line width=0.26666114429351456] (505152x) to[out=0,in=180] node[weight] {5.3\%} (5051x52);
\end{tikzpicture}
\]
Above, based on counting half of the votes, we see a very high probability that Franks will be eliminated first, followed by a likely win by Groh. (This is indeed what happened in this election.)

To get a feel for the algorithm's effectiveness, we can repeat the calculation above based on counting $x$\% of the vote for various $x\in [0,100]$. For a more realistic imitation of an election night count in progress, we will simulate a partial count of votes by choosing a random ordering of the precinct portions and then counting votes in order according to their precinct portions. Thus, as in a real election, our votes will be counted in some order determined by geography.

The winning probabilities at each value of $x$ are shown in Figure \ref{house18fig}. 

\begin{figure}
\[
\begin{tikzpicture}
\begin{axis}[domain = -.6:.6, 
xtick={0,.25,.5,.75,1},
xticklabels={$0\%$, $25\%$, $50\%$, $75\%$, $100\%$},
ytick={0,.25,.5,.75,1},
yticklabels={$0\%$, $25\%$, $50\%$, $75\%$, $100\%$},
xlabel=Percentage of vote tallied,
ylabel=Probability of win,
height=4cm,
width=\winplotwidth,
scale only axis,
xlabel near ticks,
ylabel near ticks,
no marks, thick,
legend pos=outer north east
]
\addplot  coordinates { 
(0,0.333)
(0.005,0.000) (0.010,0.127) (0.015,0.135) (0.020,0.090) (0.025,0.053) (0.030,0.079) (0.035,0.043) (0.040,0.096) (0.045,0.096) (0.050,0.091) (0.055,0.071) (0.060,0.065) (0.065,0.070) (0.070,0.045) (0.075,0.053) (0.080,0.090) (0.085,0.079) (0.090,0.054) (0.095,0.093) (0.100,0.098) (0.105,0.110) (0.110,0.107) (0.115,0.105) (0.120,0.113) (0.125,0.125) (0.130,0.106) (0.135,0.147) (0.140,0.141) (0.145,0.150) (0.150,0.138) (0.155,0.124) (0.160,0.109) (0.165,0.099) (0.170,0.065) (0.175,0.066) (0.180,0.063) (0.185,0.064) (0.190,0.054) (0.195,0.060) (0.200,0.054) (0.205,0.057) (0.210,0.056) (0.215,0.043) (0.220,0.052) (0.225,0.052) (0.230,0.055) (0.235,0.067) (0.240,0.071) (0.245,0.064) (0.250,0.052) (0.255,0.055) (0.260,0.064) (0.265,0.067) (0.270,0.056) (0.275,0.056) (0.280,0.053) (0.285,0.052) (0.290,0.047) (0.295,0.045) (0.300,0.044) (0.305,0.044) (0.310,0.040) (0.315,0.052) (0.320,0.051) (0.325,0.056) (0.330,0.054) (0.335,0.051) (0.340,0.047) (0.345,0.038) (0.350,0.036) (0.355,0.038) (0.360,0.047) (0.365,0.046) (0.370,0.039) (0.375,0.033) (0.380,0.033) (0.385,0.032) (0.390,0.033) (0.395,0.028) (0.400,0.030) (0.405,0.030) (0.410,0.032) (0.415,0.032) (0.420,0.053) (0.425,0.050) (0.430,0.048) (0.435,0.049) (0.440,0.036) (0.445,0.039) (0.450,0.039) (0.455,0.037) (0.460,0.037) (0.465,0.036) (0.470,0.035) (0.475,0.027) (0.480,0.025) (0.485,0.022) (0.490,0.016) (0.495,0.015) (0.500,0.014) (0.505,0.014) (0.510,0.013) (0.515,0.012) (0.520,0.013) (0.525,0.009) (0.530,0.006) (0.535,0.008) (0.540,0.013) (0.545,0.007) (0.550,0.008) (0.555,0.008) (0.560,0.007) (0.565,0.004) (0.570,0.003) (0.575,0.003) (0.580,0.002) (0.585,0.002) (0.590,0.002) (0.595,0.001) (0.600,0.001) (0.605,0.001) (0.610,0.001) (0.615,0.001) (0.620,0.000) (0.625,0.000) (0.630,0.000) (0.635,0.000) (0.640,0.000) (0.645,0.000) (0.650,0.000) (0.655,0.000) (0.660,0.000) (0.665,0.000) (0.670,0.000) (0.675,0.000) (0.680,0.000) (0.685,0.000) (0.690,0.000) (0.695,0.000) (0.700,0.000) (0.705,0.000) (0.710,0.000) (0.715,0.000) (0.720,0.000) (0.725,0.000) (0.730,0.000) (0.735,0.000) (0.740,0.000) (0.745,0.000) (0.750,0.000) (0.755,0.000) (0.760,0.000) (0.765,0.000) (0.770,0.000) (0.775,0.000) (0.780,0.000) (0.785,0.000) (0.790,0.000) (0.795,0.000) (0.800,-0.000) (0.805,-0.000) (0.810,-0.000) (0.815,-0.000) (0.820,-0.000) (0.825,0.000) (0.830,-0.000) (0.835,0.000) (0.840,0.000) (0.845,0.000) (0.850,-0.000) (0.855,0.000) (0.860,0.000) (0.865,0.000) (0.870,-0.000) (0.875,0.000) (0.880,0.000) (0.885,-0.000) (0.890,0.000) (0.895,-0.000) (0.900,0.000) (0.905,-0.000) (0.910,0.000) (0.915,0.000) (0.920,0.000) (0.925,0.000) (0.930,0.000) (0.935,0.000) (0.940,0.000) (0.945,0.000) (0.950,0.000) (0.955,0.000) (0.960,-0.000) (0.965,-0.000) (0.970,-0.000) (0.975,-0.000) (0.980,0.000) (0.985,0.000) (0.990,0.000) (0.995,0.000) (1.000,0.000) 
};
\addlegendentry{Franks}

\addplot  coordinates { 
(0,0.333)
(0.005,0.342) (0.010,0.307) (0.015,0.110) (0.020,0.316) (0.025,0.368) (0.030,0.283) (0.035,0.357) (0.040,0.376) (0.045,0.333) (0.050,0.363) (0.055,0.408) (0.060,0.365) (0.065,0.441) (0.070,0.394) (0.075,0.419) (0.080,0.418) (0.085,0.460) (0.090,0.446) (0.095,0.409) (0.100,0.410) (0.105,0.387) (0.110,0.387) (0.115,0.393) (0.120,0.384) (0.125,0.376) (0.130,0.393) (0.135,0.347) (0.140,0.335) (0.145,0.337) (0.150,0.316) (0.155,0.303) (0.160,0.287) (0.165,0.267) (0.170,0.241) (0.175,0.222) (0.180,0.240) (0.185,0.257) (0.190,0.262) (0.195,0.259) (0.200,0.273) (0.205,0.263) (0.210,0.282) (0.215,0.256) (0.220,0.256) (0.225,0.238) (0.230,0.241) (0.235,0.215) (0.240,0.211) (0.245,0.244) (0.250,0.238) (0.255,0.257) (0.260,0.256) (0.265,0.259) (0.270,0.289) (0.275,0.281) (0.280,0.295) (0.285,0.296) (0.290,0.319) (0.295,0.323) (0.300,0.295) (0.305,0.299) (0.310,0.307) (0.315,0.336) (0.320,0.341) (0.325,0.339) (0.330,0.340) (0.335,0.340) (0.340,0.360) (0.345,0.332) (0.350,0.323) (0.355,0.326) (0.360,0.298) (0.365,0.293) (0.370,0.273) (0.375,0.245) (0.380,0.235) (0.385,0.258) (0.390,0.251) (0.395,0.234) (0.400,0.232) (0.405,0.227) (0.410,0.221) (0.415,0.236) (0.420,0.211) (0.425,0.228) (0.430,0.228) (0.435,0.225) (0.440,0.205) (0.445,0.210) (0.450,0.210) (0.455,0.211) (0.460,0.211) (0.465,0.212) (0.470,0.212) (0.475,0.188) (0.480,0.210) (0.485,0.203) (0.490,0.183) (0.495,0.177) (0.500,0.179) (0.505,0.164) (0.510,0.193) (0.515,0.215) (0.520,0.212) (0.525,0.193) (0.530,0.173) (0.535,0.155) (0.540,0.138) (0.545,0.119) (0.550,0.113) (0.555,0.127) (0.560,0.130) (0.565,0.110) (0.570,0.110) (0.575,0.107) (0.580,0.112) (0.585,0.108) (0.590,0.120) (0.595,0.113) (0.600,0.111) (0.605,0.109) (0.610,0.086) (0.615,0.092) (0.620,0.077) (0.625,0.075) (0.630,0.087) (0.635,0.066) (0.640,0.053) (0.645,0.053) (0.650,0.053) (0.655,0.064) (0.660,0.061) (0.665,0.059) (0.670,0.061) (0.675,0.061) (0.680,0.072) (0.685,0.061) (0.690,0.052) (0.695,0.052) (0.700,0.077) (0.705,0.077) (0.710,0.079) (0.715,0.079) (0.720,0.095) (0.725,0.096) (0.730,0.096) (0.735,0.090) (0.740,0.106) (0.745,0.128) (0.750,0.103) (0.755,0.083) (0.760,0.083) (0.765,0.103) (0.770,0.101) (0.775,0.123) (0.780,0.157) (0.785,0.160) (0.790,0.160) (0.795,0.196) (0.800,0.197) (0.805,0.196) (0.810,0.194) (0.815,0.120) (0.820,0.149) (0.825,0.183) (0.830,0.178) (0.835,0.172) (0.840,0.221) (0.845,0.210) (0.850,0.241) (0.855,0.311) (0.860,0.311) (0.865,0.313) (0.870,0.239) (0.875,0.246) (0.880,0.230) (0.885,0.226) (0.890,0.121) (0.895,0.107) (0.900,0.045) (0.905,0.048) (0.910,0.022) (0.915,0.035) (0.920,0.034) (0.925,0.025) (0.930,0.041) (0.935,0.092) (0.940,0.098) (0.945,0.028) (0.950,0.042) (0.955,0.033) (0.960,0.011) (0.965,0.028) (0.970,0.034) (0.975,0.000) (0.980,0.000) (0.985,0.000) (0.990,0.000) (0.995,0.000) (1.000,0.000) 
};
\addlegendentry{Nelson}

\addplot  coordinates { 
(0,0.333)
(0.005,0.658) (0.010,0.566) (0.015,0.755) (0.020,0.594) (0.025,0.580) (0.030,0.637) (0.035,0.601) (0.040,0.529) (0.045,0.571) (0.050,0.547) (0.055,0.521) (0.060,0.569) (0.065,0.489) (0.070,0.562) (0.075,0.528) (0.080,0.492) (0.085,0.461) (0.090,0.500) (0.095,0.498) (0.100,0.492) (0.105,0.502) (0.110,0.506) (0.115,0.503) (0.120,0.503) (0.125,0.498) (0.130,0.500) (0.135,0.506) (0.140,0.524) (0.145,0.513) (0.150,0.546) (0.155,0.572) (0.160,0.604) (0.165,0.634) (0.170,0.695) (0.175,0.713) (0.180,0.697) (0.185,0.679) (0.190,0.684) (0.195,0.681) (0.200,0.673) (0.205,0.679) (0.210,0.662) (0.215,0.700) (0.220,0.693) (0.225,0.710) (0.230,0.703) (0.235,0.718) (0.240,0.718) (0.245,0.692) (0.250,0.710) (0.255,0.688) (0.260,0.680) (0.265,0.674) (0.270,0.656) (0.275,0.663) (0.280,0.652) (0.285,0.652) (0.290,0.634) (0.295,0.632) (0.300,0.661) (0.305,0.658) (0.310,0.653) (0.315,0.613) (0.320,0.607) (0.325,0.604) (0.330,0.606) (0.335,0.609) (0.340,0.593) (0.345,0.630) (0.350,0.641) (0.355,0.636) (0.360,0.655) (0.365,0.661) (0.370,0.688) (0.375,0.722) (0.380,0.732) (0.385,0.711) (0.390,0.716) (0.395,0.738) (0.400,0.738) (0.405,0.743) (0.410,0.747) (0.415,0.732) (0.420,0.736) (0.425,0.722) (0.430,0.724) (0.435,0.726) (0.440,0.759) (0.445,0.751) (0.450,0.751) (0.455,0.752) (0.460,0.752) (0.465,0.752) (0.470,0.753) (0.475,0.784) (0.480,0.765) (0.485,0.775) (0.490,0.802) (0.495,0.808) (0.500,0.807) (0.505,0.821) (0.510,0.795) (0.515,0.772) (0.520,0.776) (0.525,0.798) (0.530,0.822) (0.535,0.836) (0.540,0.850) (0.545,0.874) (0.550,0.879) (0.555,0.865) (0.560,0.863) (0.565,0.886) (0.570,0.887) (0.575,0.891) (0.580,0.886) (0.585,0.890) (0.590,0.878) (0.595,0.886) (0.600,0.888) (0.605,0.890) (0.610,0.913) (0.615,0.907) (0.620,0.923) (0.625,0.924) (0.630,0.913) (0.635,0.934) (0.640,0.946) (0.645,0.947) (0.650,0.947) (0.655,0.936) (0.660,0.939) (0.665,0.941) (0.670,0.939) (0.675,0.939) (0.680,0.928) (0.685,0.939) (0.690,0.948) (0.695,0.948) (0.700,0.923) (0.705,0.923) (0.710,0.921) (0.715,0.921) (0.720,0.905) (0.725,0.904) (0.730,0.904) (0.735,0.910) (0.740,0.894) (0.745,0.872) (0.750,0.897) (0.755,0.917) (0.760,0.917) (0.765,0.897) (0.770,0.899) (0.775,0.877) (0.780,0.843) (0.785,0.840) (0.790,0.840) (0.795,0.804) (0.800,0.803) (0.805,0.804) (0.810,0.806) (0.815,0.880) (0.820,0.851) (0.825,0.817) (0.830,0.822) (0.835,0.828) (0.840,0.779) (0.845,0.790) (0.850,0.759) (0.855,0.689) (0.860,0.689) (0.865,0.687) (0.870,0.761) (0.875,0.754) (0.880,0.770) (0.885,0.774) (0.890,0.879) (0.895,0.893) (0.900,0.955) (0.905,0.952) (0.910,0.978) (0.915,0.965) (0.920,0.966) (0.925,0.975) (0.930,0.959) (0.935,0.908) (0.940,0.902) (0.945,0.972) (0.950,0.958) (0.955,0.967) (0.960,0.989) (0.965,0.972) (0.970,0.966) (0.975,1.000) (0.980,1.000) (0.985,1.000) (0.990,1.000) (0.995,1.000) (1.000,1.000) 
};
\addlegendentry{Groh}
\end{axis}
\end{tikzpicture}
\]
\caption{Win probabilities for Alaska House 18, as votes are tallied\label{house18fig}}
\end{figure}
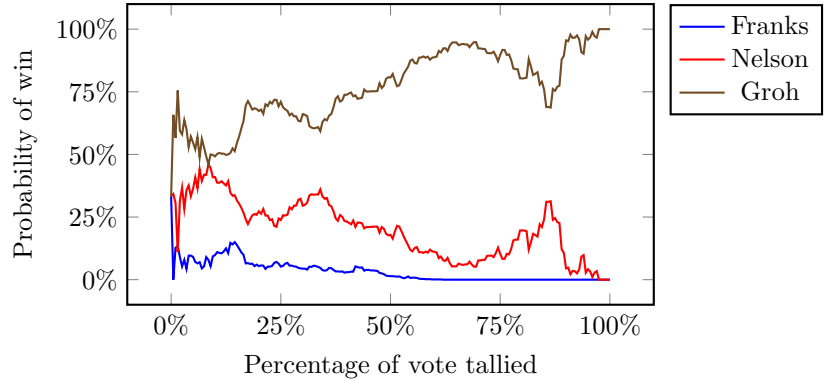

\begin{figure}
\[
\begin{tikzpicture}[scale=1]
\def\pgfplotspointmetatransformed{1000}
\begin{ternaryaxis}[
no marks,
axis line style={densely dotted},
xtick={0,.25,.50,.75,1.00},
ytick={0,.25,.50,.75,1.00},
ztick={0,.25,.50,.75,1.00},
xticklabels={$0\%$,$25\%$,$50\%$,$75\%$,$100\%$},
yticklabels={$0\%$,$25\%$,$50\%$,$75\%$,$100\%$},
zticklabels={$0\%$,$25\%$,$50\%$,$75\%$,$100\%$},
xlabel=Franks win probability,
ylabel=Nelson win probability,
zlabel=Groh win probability,
label style=sloped,
 ]
    \addplot3[mesh,thick,color=teal,
    no marks,
    ] coordinates {
(0.333,0.333,0.333)
(0.000,0.342,0.658) (0.127,0.307,0.566) (0.135,0.110,0.755) (0.090,0.316,0.594) (0.053,0.368,0.580) (0.079,0.283,0.637) (0.043,0.357,0.601) (0.096,0.376,0.529) (0.096,0.333,0.571) (0.091,0.363,0.547) (0.071,0.408,0.521) (0.065,0.365,0.569) (0.070,0.441,0.489) (0.045,0.394,0.562) (0.053,0.419,0.528) (0.090,0.418,0.492) (0.079,0.460,0.461) (0.054,0.446,0.500) (0.093,0.409,0.498) (0.098,0.410,0.492) (0.110,0.387,0.502) (0.107,0.387,0.506) (0.105,0.393,0.503) (0.113,0.384,0.503) (0.125,0.376,0.498) (0.106,0.393,0.500) (0.147,0.347,0.506) (0.141,0.335,0.524) (0.150,0.337,0.513) (0.138,0.316,0.546) (0.124,0.303,0.572) (0.109,0.287,0.604) (0.099,0.267,0.634) (0.065,0.241,0.695) (0.066,0.222,0.713) (0.063,0.240,0.697) (0.064,0.257,0.679) (0.054,0.262,0.684) (0.060,0.259,0.681) (0.054,0.273,0.673) (0.057,0.263,0.679) (0.056,0.282,0.662) (0.043,0.256,0.700) (0.052,0.256,0.693) (0.052,0.238,0.710) (0.055,0.241,0.703) (0.067,0.215,0.718) (0.071,0.211,0.718) (0.064,0.244,0.692) (0.052,0.238,0.710) (0.055,0.257,0.688) (0.064,0.256,0.680) (0.067,0.259,0.674) (0.056,0.289,0.656) (0.056,0.281,0.663) (0.053,0.295,0.652) (0.052,0.296,0.652) (0.047,0.319,0.634) (0.045,0.323,0.632) (0.044,0.295,0.661) (0.044,0.299,0.658) (0.040,0.307,0.653) (0.052,0.336,0.613) (0.051,0.341,0.607) (0.056,0.339,0.604) (0.054,0.340,0.606) (0.051,0.340,0.609) (0.047,0.360,0.593) (0.038,0.332,0.630) (0.036,0.323,0.641) (0.038,0.326,0.636) (0.047,0.298,0.655) (0.046,0.293,0.661) (0.039,0.273,0.688) (0.033,0.245,0.722) (0.033,0.235,0.732) (0.032,0.258,0.711) (0.033,0.251,0.716) (0.028,0.234,0.738) (0.030,0.232,0.738) (0.030,0.227,0.743) (0.032,0.221,0.747) (0.032,0.236,0.732) (0.053,0.211,0.736) (0.050,0.228,0.722) (0.048,0.228,0.724) (0.049,0.225,0.726) (0.036,0.205,0.759) (0.039,0.210,0.751) (0.039,0.210,0.751) (0.037,0.211,0.752) (0.037,0.211,0.752) (0.036,0.212,0.752) (0.035,0.212,0.753) (0.027,0.188,0.784) (0.025,0.210,0.765) (0.022,0.203,0.775) (0.016,0.183,0.802) (0.015,0.177,0.808) (0.014,0.179,0.807) (0.014,0.164,0.821) (0.013,0.193,0.795) (0.012,0.215,0.772) (0.013,0.212,0.776) (0.009,0.193,0.798) (0.006,0.173,0.822) (0.008,0.155,0.836) (0.013,0.138,0.850) (0.007,0.119,0.874) (0.008,0.113,0.879) (0.008,0.127,0.865) (0.007,0.130,0.863) (0.004,0.110,0.886) (0.003,0.110,0.887) (0.003,0.107,0.891) (0.002,0.112,0.886) (0.002,0.108,0.890) (0.002,0.120,0.878) (0.001,0.113,0.886) (0.001,0.111,0.888) (0.001,0.109,0.890) (0.001,0.086,0.913) (0.001,0.092,0.907) (0.000,0.077,0.923) (0.000,0.075,0.924) (0.000,0.087,0.913) (0.000,0.066,0.934) (0.000,0.053,0.946) (0.000,0.053,0.947) (0.000,0.053,0.947) (0.000,0.064,0.936) (0.000,0.061,0.939) (0.000,0.059,0.941) (0.000,0.061,0.939) (0.000,0.061,0.939) (0.000,0.072,0.928) (0.000,0.061,0.939) (0.000,0.052,0.948) (0.000,0.052,0.948) (0.000,0.077,0.923) (0.000,0.077,0.923) (0.000,0.079,0.921) (0.000,0.079,0.921) (0.000,0.095,0.905) (0.000,0.096,0.904) (0.000,0.096,0.904) (0.000,0.090,0.910) (0.000,0.106,0.894) (0.000,0.128,0.872) (0.000,0.103,0.897) (0.000,0.083,0.917) (0.000,0.083,0.917) (0.000,0.103,0.897) (0.000,0.101,0.899) (0.000,0.123,0.877) (0.000,0.157,0.843) (0.000,0.160,0.840) (0.000,0.160,0.840) (0.000,0.196,0.804) (-0.000,0.197,0.803) (-0.000,0.196,0.804) (-0.000,0.194,0.806) (-0.000,0.120,0.880) (-0.000,0.149,0.851) (0.000,0.183,0.817) (-0.000,0.178,0.822) (0.000,0.172,0.828) (0.000,0.221,0.779) (0.000,0.210,0.790) (-0.000,0.241,0.759) (0.000,0.311,0.689) (0.000,0.311,0.689) (0.000,0.313,0.687) (-0.000,0.239,0.761) (0.000,0.246,0.754) (0.000,0.230,0.770) (-0.000,0.226,0.774) (0.000,0.121,0.879) (-0.000,0.107,0.893) (0.000,0.045,0.955) (-0.000,0.048,0.952) (0.000,0.022,0.978) (0.000,0.035,0.965) (0.000,0.034,0.966) (0.000,0.025,0.975) (0.000,0.041,0.959) (0.000,0.092,0.908) (0.000,0.098,0.902) (0.000,0.028,0.972) (0.000,0.042,0.958) (0.000,0.033,0.967) (-0.000,0.011,0.989) (-0.000,0.028,0.972) (-0.000,0.034,0.966) (-0.000,0.000,1.000) (0.000,0.000,1.000) (0.000,0.000,1.000) (0.000,0.000,1.000) (0.000,0.000,1.000) (0.000,0.000,1.000) 
    }; 
\end{ternaryaxis}
\end{tikzpicture}
\]
\caption{Ternary plot of data from Figure \ref{house18fig}\label{house18ternaryfig}}
\end{figure}
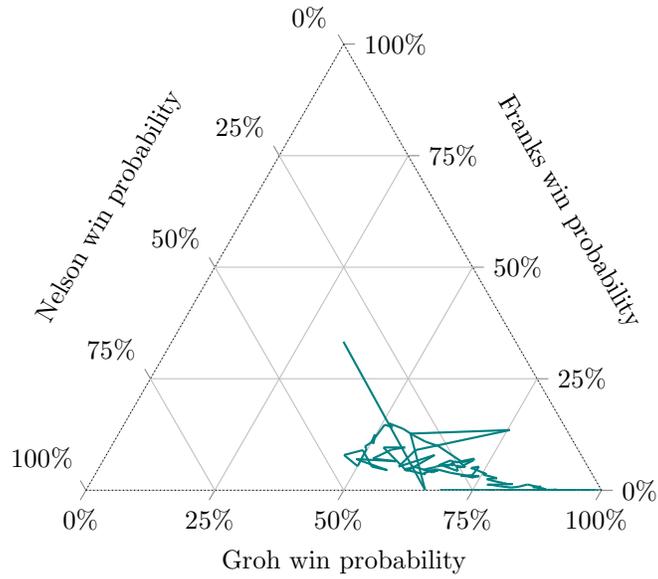

A visualization like Figure \ref{house18fig} would be a convenient way to display live predictions of an in-progress election on election night. Early during the counting process (on the left side of the diagram), the predictions are less specific, and as more and more votes are counted, it becomes clearer who the eventual winner will be. Since there are exactly 3 candidates, we can also visualize the same data in the ternary plot of Figure \ref{house18ternaryfig}.


\end{exl}

\begin{exl}{\bf 2022 US Senator from Alaska}\label{ussenateexl}

The 2022 ballot for US Senator from Alaska featured Republican incumbent and eventual winner Lisa Murkowski, along with fellow Republican Kelly Tshibaka, as the main contenders. Democrat Pat Chesbro also appeared on the ballot, as did Buzz Kelley, a Republican who had suspended his campaign and officially endorsed Tshibaka two months before the election. Performing the same analysis as above gives Figure \ref{ussenatefig}.

\newcommand{\murkowskicolor}{violet}
\newcommand{\tshibakacolor}{orange}

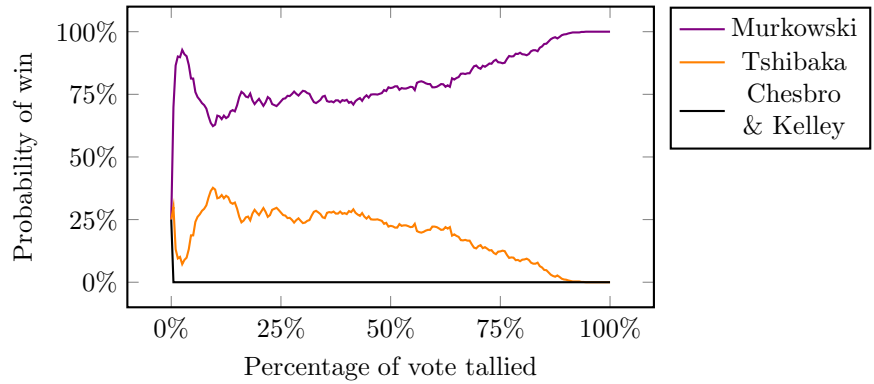
\begin{figure}
\[
\begin{tikzpicture} 
\begin{axis}[domain = 0:1, 
xtick={0,.25,.5,.75,1},
xticklabels={$0\%$, $25\%$, $50\%$, $75\%$, $100\%$},
ytick={0,.25,.5,.75,1},
yticklabels={$0\%$, $25\%$, $50\%$, $75\%$, $100\%$},
xlabel=Percentage of vote tallied,
ylabel=Probability of win,
height=4cm,
width=\winplotwidth,
scale only axis,
xlabel near ticks,
ylabel near ticks,
no marks, thick,
legend pos=outer north east,
legend style={cells={align=center}}
]
\addplot+[color=\murkowskicolor]  coordinates { 
(0,0.25)
(0.005,0.698) (0.010,0.865) (0.015,0.901) (0.020,0.898) (0.025,0.927) (0.030,0.910) (0.035,0.902) (0.040,0.867) (0.045,0.813) (0.050,0.813) (0.055,0.759) (0.060,0.739) (0.065,0.728) (0.070,0.715) (0.075,0.707) (0.080,0.693) (0.085,0.662) (0.090,0.636) (0.095,0.623) (0.100,0.629) (0.105,0.665) (0.110,0.662) (0.115,0.651) (0.120,0.665) (0.125,0.655) (0.130,0.661) (0.135,0.682) (0.140,0.687) (0.145,0.682) (0.150,0.706) (0.155,0.739) (0.160,0.760) (0.165,0.753) (0.170,0.741) (0.175,0.738) (0.180,0.753) (0.185,0.726) (0.190,0.711) (0.195,0.722) (0.200,0.732) (0.205,0.718) (0.210,0.704) (0.215,0.718) (0.220,0.739) (0.225,0.732) (0.230,0.711) (0.235,0.707) (0.240,0.703) (0.245,0.712) (0.250,0.722) (0.255,0.732) (0.260,0.734) (0.265,0.744) (0.270,0.743) (0.275,0.751) (0.280,0.761) (0.285,0.752) (0.290,0.745) (0.295,0.755) (0.300,0.764) (0.305,0.762) (0.310,0.755) (0.315,0.752) (0.320,0.734) (0.325,0.719) (0.330,0.715) (0.335,0.721) (0.340,0.728) (0.345,0.741) (0.350,0.744) (0.355,0.723) (0.360,0.720) (0.365,0.722) (0.370,0.718) (0.375,0.726) (0.380,0.726) (0.385,0.718) (0.390,0.727) (0.395,0.724) (0.400,0.728) (0.405,0.716) (0.410,0.720) (0.415,0.709) (0.420,0.723) (0.425,0.730) (0.430,0.735) (0.435,0.724) (0.440,0.738) (0.445,0.746) (0.450,0.737) (0.455,0.749) (0.460,0.750) (0.465,0.750) (0.470,0.749) (0.475,0.752) (0.480,0.755) (0.485,0.767) (0.490,0.763) (0.495,0.778) (0.500,0.775) (0.505,0.777) (0.510,0.782) (0.515,0.782) (0.520,0.768) (0.525,0.775) (0.530,0.773) (0.535,0.773) (0.540,0.776) (0.545,0.780) (0.550,0.779) (0.555,0.767) (0.560,0.796) (0.565,0.800) (0.570,0.802) (0.575,0.799) (0.580,0.794) (0.585,0.791) (0.590,0.791) (0.595,0.778) (0.600,0.778) (0.605,0.780) (0.610,0.785) (0.615,0.791) (0.620,0.790) (0.625,0.780) (0.630,0.786) (0.635,0.780) (0.640,0.814) (0.645,0.809) (0.650,0.815) (0.655,0.819) (0.660,0.832) (0.665,0.833) (0.670,0.831) (0.675,0.834) (0.680,0.834) (0.685,0.849) (0.690,0.862) (0.695,0.865) (0.700,0.856) (0.705,0.852) (0.710,0.864) (0.715,0.860) (0.720,0.866) (0.725,0.873) (0.730,0.872) (0.735,0.883) (0.740,0.888) (0.745,0.878) (0.750,0.877) (0.755,0.874) (0.760,0.876) (0.765,0.891) (0.770,0.903) (0.775,0.901) (0.780,0.902) (0.785,0.912) (0.790,0.911) (0.795,0.916) (0.800,0.910) (0.805,0.909) (0.810,0.906) (0.815,0.911) (0.820,0.922) (0.825,0.926) (0.830,0.926) (0.835,0.922) (0.840,0.936) (0.845,0.941) (0.850,0.951) (0.855,0.952) (0.860,0.961) (0.865,0.970) (0.870,0.975) (0.875,0.978) (0.880,0.973) (0.885,0.979) (0.890,0.986) (0.895,0.989) (0.900,0.990) (0.905,0.993) (0.910,0.995) (0.915,0.997) (0.920,0.997) (0.925,0.997) (0.930,0.997) (0.935,0.999) (0.940,0.999) (0.945,1.000) (0.950,1.000) (0.955,1.000) (0.960,1.000) (0.965,1.000) (0.970,1.000) (0.975,1.000) (0.980,1.000) (0.985,1.000) (0.990,1.000) (0.995,1.000) (1.000,1.000) 
};
\addlegendentry{Murkowski}

\addplot+[color=\tshibakacolor] coordinates { 
(0,0.25)
(0.005,0.300) (0.010,0.132) (0.015,0.095) (0.020,0.101) (0.025,0.072) (0.030,0.089) (0.035,0.098) (0.040,0.132) (0.045,0.187) (0.050,0.187) (0.055,0.241) (0.060,0.261) (0.065,0.271) (0.070,0.285) (0.075,0.293) (0.080,0.307) (0.085,0.338) (0.090,0.364) (0.095,0.377) (0.100,0.370) (0.105,0.335) (0.110,0.338) (0.115,0.348) (0.120,0.335) (0.125,0.345) (0.130,0.339) (0.135,0.318) (0.140,0.313) (0.145,0.318) (0.150,0.294) (0.155,0.261) (0.160,0.239) (0.165,0.246) (0.170,0.259) (0.175,0.262) (0.180,0.247) (0.185,0.274) (0.190,0.289) (0.195,0.278) (0.200,0.268) (0.205,0.282) (0.210,0.296) (0.215,0.281) (0.220,0.261) (0.225,0.268) (0.230,0.289) (0.235,0.293) (0.240,0.297) (0.245,0.288) (0.250,0.278) (0.255,0.268) (0.260,0.266) (0.265,0.256) (0.270,0.257) (0.275,0.249) (0.280,0.238) (0.285,0.248) (0.290,0.255) (0.295,0.245) (0.300,0.236) (0.305,0.238) (0.310,0.245) (0.315,0.248) (0.320,0.266) (0.325,0.281) (0.330,0.285) (0.335,0.279) (0.340,0.272) (0.345,0.259) (0.350,0.256) (0.355,0.277) (0.360,0.280) (0.365,0.278) (0.370,0.282) (0.375,0.274) (0.380,0.274) (0.385,0.282) (0.390,0.273) (0.395,0.276) (0.400,0.272) (0.405,0.284) (0.410,0.280) (0.415,0.291) (0.420,0.277) (0.425,0.270) (0.430,0.265) (0.435,0.276) (0.440,0.262) (0.445,0.254) (0.450,0.263) (0.455,0.251) (0.460,0.250) (0.465,0.250) (0.470,0.251) (0.475,0.248) (0.480,0.245) (0.485,0.233) (0.490,0.237) (0.495,0.222) (0.500,0.225) (0.505,0.223) (0.510,0.218) (0.515,0.218) (0.520,0.232) (0.525,0.225) (0.530,0.227) (0.535,0.227) (0.540,0.224) (0.545,0.220) (0.550,0.221) (0.555,0.233) (0.560,0.204) (0.565,0.200) (0.570,0.198) (0.575,0.201) (0.580,0.206) (0.585,0.209) (0.590,0.209) (0.595,0.222) (0.600,0.222) (0.605,0.220) (0.610,0.215) (0.615,0.209) (0.620,0.210) (0.625,0.220) (0.630,0.214) (0.635,0.220) (0.640,0.186) (0.645,0.191) (0.650,0.185) (0.655,0.181) (0.660,0.168) (0.665,0.167) (0.670,0.169) (0.675,0.166) (0.680,0.166) (0.685,0.151) (0.690,0.138) (0.695,0.135) (0.700,0.144) (0.705,0.148) (0.710,0.136) (0.715,0.140) (0.720,0.134) (0.725,0.127) (0.730,0.128) (0.735,0.117) (0.740,0.112) (0.745,0.122) (0.750,0.123) (0.755,0.126) (0.760,0.124) (0.765,0.109) (0.770,0.097) (0.775,0.099) (0.780,0.098) (0.785,0.088) (0.790,0.089) (0.795,0.084) (0.800,0.090) (0.805,0.091) (0.810,0.094) (0.815,0.089) (0.820,0.078) (0.825,0.074) (0.830,0.074) (0.835,0.078) (0.840,0.064) (0.845,0.059) (0.850,0.049) (0.855,0.048) (0.860,0.039) (0.865,0.030) (0.870,0.025) (0.875,0.022) (0.880,0.027) (0.885,0.021) (0.890,0.014) (0.895,0.011) (0.900,0.010) (0.905,0.007) (0.910,0.005) (0.915,0.003) (0.920,0.003) (0.925,0.003) (0.930,0.003) (0.935,0.001) (0.940,0.001) (0.945,0.000) (0.950,0.000) (0.955,0.000) (0.960,0.000) (0.965,0.000) (0.970,0.000) (0.975,0.000) (0.980,0.000) (0.985,0.000) (0.990,0.000) (0.995,0.000) (1.000,0.000)  

};
\addlegendentry{Tshibaka}

\addplot+[color=black] coordinates { 
(0,0.25)
(0.005,0) (1.000,0.000) 
};
\addlegendentry{Chesbro \\ \& Kelley}

\end{axis}
\end{tikzpicture}
\]
\caption{Win probabilities for Alaska US Senate race, as votes are tallied\label{ussenatefig}}
\end{figure}

Early in the counting in Figure \ref{ussenatefig} we see some significant possibility of a win by Tshibaka, but as votes are counted it becomes increasingly clear that Murkowski will win. The probabilities of wins by Kelly and Chesbro never rise above $0.1\%$.

The shape of the curves in Figures \ref{house18fig} and \ref{ussenatefig} is heavily dependent on the order in which the precinct portions are tabulated. Charts resulting from three other orderings of the precinct portions in the US Senate race are given in Figure \ref{geographicfig}.

\begin{figure}
\triplechart{\[ \includegraphics[width=5in]{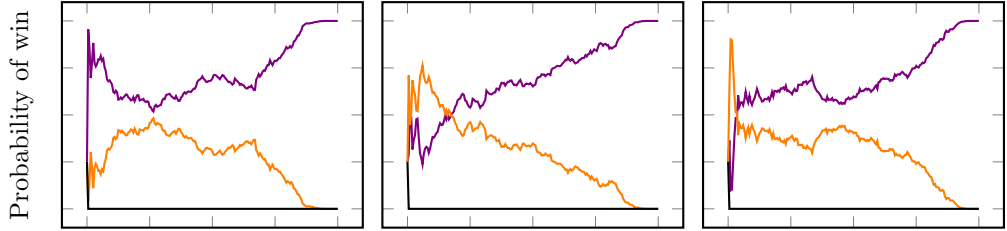} \]}{
\[
\begin{tikzpicture} 
\begin{axis}[domain = 0:1, 
xtick={0,.25,.5,.75,1},
xticklabels={},
ytick={0,.25,.5,.75,1},
yticklabels={},
xlabel=,
ylabel=Probability of win,
height=3cm,
width=\tiledchartwidth,
scale only axis,
xlabel near ticks,
ylabel near ticks,
no marks, thick,
legend pos=outer north east
]
\addplot+[color=\murkowskicolor]  coordinates { 
(0,0.25)
(0.005,0.955) (0.010,0.856) (0.015,0.696) (0.020,0.778) (0.025,0.883) (0.030,0.778) (0.035,0.758) (0.040,0.770) (0.045,0.795) (0.050,0.794) (0.055,0.814) (0.060,0.788) (0.065,0.798) (0.070,0.754) (0.075,0.733) (0.080,0.688) (0.085,0.679) (0.090,0.684) (0.095,0.673) (0.100,0.661) (0.105,0.622) (0.110,0.616) (0.115,0.629) (0.120,0.621) (0.125,0.581) (0.130,0.584) (0.135,0.593) (0.140,0.584) (0.145,0.593) (0.150,0.599) (0.155,0.609) (0.160,0.607) (0.165,0.593) (0.170,0.588) (0.175,0.577) (0.180,0.580) (0.185,0.594) (0.190,0.573) (0.195,0.590) (0.200,0.604) (0.205,0.605) (0.210,0.577) (0.215,0.581) (0.220,0.581) (0.225,0.585) (0.230,0.572) (0.235,0.569) (0.240,0.573) (0.245,0.569) (0.250,0.543) (0.255,0.540) (0.260,0.525) (0.265,0.518) (0.270,0.546) (0.275,0.553) (0.280,0.543) (0.285,0.547) (0.290,0.551) (0.295,0.553) (0.300,0.576) (0.305,0.586) (0.310,0.584) (0.315,0.604) (0.320,0.608) (0.325,0.610) (0.330,0.612) (0.335,0.601) (0.340,0.596) (0.345,0.589) (0.350,0.586) (0.355,0.600) (0.360,0.586) (0.365,0.574) (0.370,0.575) (0.375,0.592) (0.380,0.599) (0.385,0.614) (0.390,0.620) (0.395,0.613) (0.400,0.621) (0.405,0.625) (0.410,0.623) (0.415,0.622) (0.420,0.631) (0.425,0.639) (0.430,0.639) (0.435,0.651) (0.440,0.676) (0.445,0.685) (0.450,0.688) (0.455,0.694) (0.460,0.705) (0.465,0.693) (0.470,0.686) (0.475,0.676) (0.480,0.675) (0.485,0.681) (0.490,0.680) (0.495,0.684) (0.500,0.693) (0.505,0.711) (0.510,0.713) (0.515,0.710) (0.520,0.703) (0.525,0.697) (0.530,0.690) (0.535,0.701) (0.540,0.712) (0.545,0.709) (0.550,0.698) (0.555,0.691) (0.560,0.682) (0.565,0.674) (0.570,0.676) (0.575,0.669) (0.580,0.661) (0.585,0.667) (0.590,0.693) (0.595,0.685) (0.600,0.679) (0.605,0.672) (0.610,0.660) (0.615,0.665) (0.620,0.659) (0.625,0.657) (0.630,0.647) (0.635,0.641) (0.640,0.641) (0.645,0.649) (0.650,0.647) (0.655,0.656) (0.660,0.645) (0.665,0.633) (0.670,0.633) (0.675,0.688) (0.680,0.700) (0.685,0.717) (0.690,0.704) (0.695,0.713) (0.700,0.729) (0.705,0.737) (0.710,0.735) (0.715,0.752) (0.720,0.765) (0.725,0.765) (0.730,0.791) (0.735,0.776) (0.740,0.765) (0.745,0.781) (0.750,0.789) (0.755,0.787) (0.760,0.792) (0.765,0.802) (0.770,0.802) (0.775,0.814) (0.780,0.823) (0.785,0.833) (0.790,0.846) (0.795,0.861) (0.800,0.853) (0.805,0.857) (0.810,0.866) (0.815,0.858) (0.820,0.868) (0.825,0.891) (0.830,0.895) (0.835,0.907) (0.840,0.922) (0.845,0.923) (0.850,0.938) (0.855,0.950) (0.860,0.972) (0.865,0.977) (0.870,0.983) (0.875,0.987) (0.880,0.987) (0.885,0.986) (0.890,0.988) (0.895,0.989) (0.900,0.990) (0.905,0.993) (0.910,0.994) (0.915,0.995) (0.920,0.996) (0.925,0.997) (0.930,0.999) (0.935,0.999) (0.940,0.999) (0.945,1.000) (0.950,1.000) (0.955,1.000) (0.960,1.000) (0.965,1.000) (0.970,1.000) (0.975,1.000) (0.980,1.000) (0.985,1.000) (0.990,1.000) (0.995,1.000) (1.000,1.000) 
};

\addplot+[color=\tshibakacolor] coordinates { 
(0,0.25)
(0.005,0.039) (0.010,0.141) (0.015,0.303) (0.020,0.219) (0.025,0.110) (0.030,0.217) (0.035,0.240) (0.040,0.227) (0.045,0.203) (0.050,0.204) (0.055,0.184) (0.060,0.210) (0.065,0.200) (0.070,0.244) (0.075,0.266) (0.080,0.311) (0.085,0.320) (0.090,0.315) (0.095,0.326) (0.100,0.338) (0.105,0.378) (0.110,0.383) (0.115,0.371) (0.120,0.378) (0.125,0.419) (0.130,0.416) (0.135,0.407) (0.140,0.416) (0.145,0.407) (0.150,0.401) (0.155,0.391) (0.160,0.393) (0.165,0.407) (0.170,0.411) (0.175,0.423) (0.180,0.419) (0.185,0.406) (0.190,0.427) (0.195,0.410) (0.200,0.396) (0.205,0.395) (0.210,0.423) (0.215,0.419) (0.220,0.419) (0.225,0.415) (0.230,0.428) (0.235,0.431) (0.240,0.426) (0.245,0.431) (0.250,0.457) (0.255,0.460) (0.260,0.475) (0.265,0.482) (0.270,0.454) (0.275,0.447) (0.280,0.457) (0.285,0.453) (0.290,0.449) (0.295,0.447) (0.300,0.424) (0.305,0.414) (0.310,0.416) (0.315,0.396) (0.320,0.392) (0.325,0.390) (0.330,0.388) (0.335,0.399) (0.340,0.404) (0.345,0.411) (0.350,0.414) (0.355,0.400) (0.360,0.414) (0.365,0.426) (0.370,0.425) (0.375,0.408) (0.380,0.401) (0.385,0.386) (0.390,0.380) (0.395,0.387) (0.400,0.379) (0.405,0.375) (0.410,0.377) (0.415,0.378) (0.420,0.369) (0.425,0.361) (0.430,0.361) (0.435,0.349) (0.440,0.324) (0.445,0.315) (0.450,0.312) (0.455,0.306) (0.460,0.295) (0.465,0.307) (0.470,0.314) (0.475,0.324) (0.480,0.325) (0.485,0.319) (0.490,0.320) (0.495,0.316) (0.500,0.307) (0.505,0.289) (0.510,0.287) (0.515,0.290) (0.520,0.297) (0.525,0.303) (0.530,0.310) (0.535,0.299) (0.540,0.288) (0.545,0.291) (0.550,0.302) (0.555,0.309) (0.560,0.318) (0.565,0.326) (0.570,0.324) (0.575,0.331) (0.580,0.339) (0.585,0.333) (0.590,0.307) (0.595,0.315) (0.600,0.321) (0.605,0.328) (0.610,0.340) (0.615,0.335) (0.620,0.341) (0.625,0.343) (0.630,0.353) (0.635,0.359) (0.640,0.359) (0.645,0.351) (0.650,0.353) (0.655,0.344) (0.660,0.355) (0.665,0.367) (0.670,0.367) (0.675,0.312) (0.680,0.300) (0.685,0.283) (0.690,0.296) (0.695,0.287) (0.700,0.271) (0.705,0.263) (0.710,0.265) (0.715,0.248) (0.720,0.235) (0.725,0.235) (0.730,0.209) (0.735,0.224) (0.740,0.235) (0.745,0.219) (0.750,0.211) (0.755,0.213) (0.760,0.208) (0.765,0.198) (0.770,0.198) (0.775,0.186) (0.780,0.177) (0.785,0.167) (0.790,0.154) (0.795,0.139) (0.800,0.147) (0.805,0.143) (0.810,0.134) (0.815,0.142) (0.820,0.132) (0.825,0.109) (0.830,0.105) (0.835,0.093) (0.840,0.078) (0.845,0.077) (0.850,0.062) (0.855,0.050) (0.860,0.028) (0.865,0.023) (0.870,0.017) (0.875,0.013) (0.880,0.013) (0.885,0.014) (0.890,0.012) (0.895,0.011) (0.900,0.010) (0.905,0.007) (0.910,0.006) (0.915,0.005) (0.920,0.004) (0.925,0.003) (0.930,0.001) (0.935,0.001) (0.940,0.001) (0.945,0.000) (0.950,0.000) (0.955,0.000) (0.960,0.000) (0.965,0.000) (0.970,0.000) (0.975,0.000) (0.980,0.000) (0.985,0.000) (0.990,0.000) (0.995,0.000) (1.000,0.000) 
};

\addplot+[color=black] coordinates { 
(0,0.25)
(0.005,0) (1.000,0.000) 
};

\end{axis}
\end{tikzpicture}
\begin{tikzpicture} 
\begin{axis}[domain = 0:1, 
xtick={0,.25,.5,.75,1},
xticklabels={},
ytick={0,.25,.5,.75,1},
yticklabels={},
xlabel=,
ylabel=,
height=3cm,
width=\tiledchartwidth,
scale only axis,
xlabel near ticks,
ylabel near ticks,
no marks, thick,
legend pos=outer north east
]
\addplot+[color=\murkowskicolor]  coordinates { 
(0,0.25)
(0.005,0.290) (0.010,0.553) (0.015,0.486) (0.020,0.316) (0.025,0.367) (0.030,0.392) (0.035,0.476) (0.040,0.480) (0.045,0.380) (0.050,0.296) (0.055,0.270) (0.060,0.237) (0.065,0.292) (0.070,0.320) (0.075,0.330) (0.080,0.326) (0.085,0.298) (0.090,0.319) (0.095,0.353) (0.100,0.382) (0.105,0.379) (0.110,0.389) (0.115,0.403) (0.120,0.408) (0.125,0.437) (0.130,0.456) (0.135,0.447) (0.140,0.447) (0.145,0.464) (0.150,0.492) (0.155,0.475) (0.160,0.499) (0.165,0.476) (0.170,0.486) (0.175,0.506) (0.180,0.509) (0.185,0.519) (0.190,0.543) (0.195,0.553) (0.200,0.578) (0.205,0.584) (0.210,0.587) (0.215,0.604) (0.220,0.612) (0.225,0.606) (0.230,0.577) (0.235,0.573) (0.240,0.555) (0.245,0.544) (0.250,0.539) (0.255,0.551) (0.260,0.586) (0.265,0.582) (0.270,0.572) (0.275,0.571) (0.280,0.571) (0.285,0.555) (0.290,0.538) (0.295,0.548) (0.300,0.551) (0.305,0.566) (0.310,0.578) (0.315,0.629) (0.320,0.620) (0.325,0.614) (0.330,0.610) (0.335,0.609) (0.340,0.614) (0.345,0.619) (0.350,0.620) (0.355,0.640) (0.360,0.639) (0.365,0.633) (0.370,0.655) (0.375,0.671) (0.380,0.668) (0.385,0.670) (0.390,0.656) (0.395,0.666) (0.400,0.676) (0.405,0.676) (0.410,0.683) (0.415,0.683) (0.420,0.668) (0.425,0.659) (0.430,0.661) (0.435,0.657) (0.440,0.667) (0.445,0.678) (0.450,0.686) (0.455,0.670) (0.460,0.653) (0.465,0.653) (0.470,0.663) (0.475,0.667) (0.480,0.664) (0.485,0.663) (0.490,0.669) (0.495,0.661) (0.500,0.652) (0.505,0.658) (0.510,0.661) (0.515,0.688) (0.520,0.700) (0.525,0.700) (0.530,0.704) (0.535,0.716) (0.540,0.714) (0.545,0.715) (0.550,0.706) (0.555,0.713) (0.560,0.718) (0.565,0.716) (0.570,0.712) (0.575,0.742) (0.580,0.748) (0.585,0.749) (0.590,0.754) (0.595,0.759) (0.600,0.761) (0.605,0.759) (0.610,0.760) (0.615,0.767) (0.620,0.776) (0.625,0.782) (0.630,0.793) (0.635,0.791) (0.640,0.808) (0.645,0.793) (0.650,0.788) (0.655,0.797) (0.660,0.793) (0.665,0.780) (0.670,0.774) (0.675,0.778) (0.680,0.781) (0.685,0.792) (0.690,0.794) (0.695,0.802) (0.700,0.802) (0.705,0.804) (0.710,0.818) (0.715,0.811) (0.720,0.822) (0.725,0.825) (0.730,0.841) (0.735,0.828) (0.740,0.839) (0.745,0.849) (0.750,0.844) (0.755,0.840) (0.760,0.861) (0.765,0.878) (0.770,0.882) (0.775,0.878) (0.780,0.868) (0.785,0.876) (0.790,0.878) (0.795,0.871) (0.800,0.872) (0.805,0.853) (0.810,0.859) (0.815,0.853) (0.820,0.858) (0.825,0.864) (0.830,0.865) (0.835,0.884) (0.840,0.900) (0.845,0.909) (0.850,0.918) (0.855,0.923) (0.860,0.946) (0.865,0.956) (0.870,0.965) (0.875,0.967) (0.880,0.967) (0.885,0.976) (0.890,0.978) (0.895,0.980) (0.900,0.983) (0.905,0.991) (0.910,0.993) (0.915,0.993) (0.920,0.993) (0.925,0.996) (0.930,0.997) (0.935,0.999) (0.940,0.999) (0.945,0.999) (0.950,1.000) (0.955,1.000) (0.960,1.000) (0.965,1.000) (0.970,1.000) (0.975,1.000) (0.980,1.000) (0.985,1.000) (0.990,1.000) (0.995,1.000) (1.000,1.000)  
};

\addplot+[color=\tshibakacolor] coordinates { 
(0,0.25)
(0.005,0.710) (0.010,0.446) (0.015,0.514) (0.020,0.684) (0.025,0.633) (0.030,0.608) (0.035,0.524) (0.040,0.520) (0.045,0.620) (0.050,0.704) (0.055,0.730) (0.060,0.763) (0.065,0.708) (0.070,0.680) (0.075,0.670) (0.080,0.674) (0.085,0.702) (0.090,0.681) (0.095,0.646) (0.100,0.618) (0.105,0.621) (0.110,0.611) (0.115,0.597) (0.120,0.592) (0.125,0.563) (0.130,0.544) (0.135,0.552) (0.140,0.553) (0.145,0.536) (0.150,0.508) (0.155,0.525) (0.160,0.501) (0.165,0.524) (0.170,0.514) (0.175,0.494) (0.180,0.491) (0.185,0.481) (0.190,0.457) (0.195,0.447) (0.200,0.422) (0.205,0.416) (0.210,0.413) (0.215,0.396) (0.220,0.388) (0.225,0.394) (0.230,0.422) (0.235,0.427) (0.240,0.445) (0.245,0.456) (0.250,0.461) (0.255,0.449) (0.260,0.414) (0.265,0.418) (0.270,0.428) (0.275,0.429) (0.280,0.429) (0.285,0.445) (0.290,0.462) (0.295,0.452) (0.300,0.449) (0.305,0.434) (0.310,0.422) (0.315,0.371) (0.320,0.380) (0.325,0.386) (0.330,0.390) (0.335,0.391) (0.340,0.386) (0.345,0.381) (0.350,0.380) (0.355,0.360) (0.360,0.361) (0.365,0.367) (0.370,0.345) (0.375,0.329) (0.380,0.332) (0.385,0.330) (0.390,0.344) (0.395,0.334) (0.400,0.324) (0.405,0.324) (0.410,0.317) (0.415,0.317) (0.420,0.332) (0.425,0.341) (0.430,0.339) (0.435,0.343) (0.440,0.333) (0.445,0.322) (0.450,0.314) (0.455,0.330) (0.460,0.347) (0.465,0.347) (0.470,0.337) (0.475,0.333) (0.480,0.336) (0.485,0.337) (0.490,0.331) (0.495,0.339) (0.500,0.348) (0.505,0.342) (0.510,0.339) (0.515,0.312) (0.520,0.300) (0.525,0.300) (0.530,0.296) (0.535,0.284) (0.540,0.286) (0.545,0.285) (0.550,0.294) (0.555,0.287) (0.560,0.282) (0.565,0.284) (0.570,0.288) (0.575,0.258) (0.580,0.252) (0.585,0.251) (0.590,0.246) (0.595,0.241) (0.600,0.239) (0.605,0.241) (0.610,0.240) (0.615,0.233) (0.620,0.224) (0.625,0.218) (0.630,0.207) (0.635,0.209) (0.640,0.192) (0.645,0.207) (0.650,0.212) (0.655,0.203) (0.660,0.207) (0.665,0.220) (0.670,0.226) (0.675,0.222) (0.680,0.219) (0.685,0.208) (0.690,0.206) (0.695,0.198) (0.700,0.198) (0.705,0.196) (0.710,0.182) (0.715,0.189) (0.720,0.178) (0.725,0.175) (0.730,0.159) (0.735,0.172) (0.740,0.161) (0.745,0.151) (0.750,0.156) (0.755,0.160) (0.760,0.139) (0.765,0.122) (0.770,0.118) (0.775,0.122) (0.780,0.132) (0.785,0.124) (0.790,0.122) (0.795,0.129) (0.800,0.128) (0.805,0.147) (0.810,0.141) (0.815,0.147) (0.820,0.142) (0.825,0.136) (0.830,0.135) (0.835,0.116) (0.840,0.100) (0.845,0.091) (0.850,0.082) (0.855,0.077) (0.860,0.054) (0.865,0.044) (0.870,0.035) (0.875,0.033) (0.880,0.033) (0.885,0.024) (0.890,0.022) (0.895,0.020) (0.900,0.017) (0.905,0.009) (0.910,0.007) (0.915,0.007) (0.920,0.007) (0.925,0.004) (0.930,0.003) (0.935,0.001) (0.940,0.001) (0.945,0.001) (0.950,0.000) (0.955,0.000) (0.960,0.000) (0.965,0.000) (0.970,0.000) (0.975,0.000) (0.980,0.000) (0.985,0.000) (0.990,0.000) (0.995,0.000) (1.000,0.000) 
};

\addplot+[color=black] coordinates { 
(0,0.25)
(0.005,0) (1.000,0.000) 
};
\end{axis}
\end{tikzpicture}
\begin{tikzpicture} 
\begin{axis}[domain = 0:1, 
xtick={0,.25,.5,.75,1},
xticklabels={},
ytick={0,.25,.5,.75,1},
yticklabels={},
xlabel=,
ylabel=,
height=3cm,
width=\tiledchartwidth,
scale only axis,
xlabel near ticks,
ylabel near ticks,
no marks, thick,
legend pos=outer north east,
legend style={cells={align=center}}
]
\addplot+[color=\murkowskicolor]  coordinates { 
(0,0.25)
(0.005,0.364) (0.010,0.098) (0.015,0.103) (0.020,0.246) (0.025,0.384) (0.030,0.483) (0.035,0.482) (0.040,0.604) (0.045,0.546) (0.050,0.566) (0.055,0.567) (0.060,0.552) (0.065,0.582) (0.070,0.623) (0.075,0.586) (0.080,0.566) (0.085,0.602) (0.090,0.570) (0.095,0.553) (0.100,0.574) (0.105,0.588) (0.110,0.611) (0.115,0.639) (0.120,0.620) (0.125,0.597) (0.130,0.589) (0.135,0.571) (0.140,0.573) (0.145,0.584) (0.150,0.584) (0.155,0.583) (0.160,0.592) (0.165,0.577) (0.170,0.597) (0.175,0.622) (0.180,0.598) (0.185,0.628) (0.190,0.627) (0.195,0.622) (0.200,0.630) (0.205,0.623) (0.210,0.641) (0.215,0.640) (0.220,0.646) (0.225,0.636) (0.230,0.655) (0.235,0.628) (0.240,0.630) (0.245,0.623) (0.250,0.619) (0.255,0.610) (0.260,0.628) (0.265,0.638) (0.270,0.626) (0.275,0.628) (0.280,0.661) (0.285,0.660) (0.290,0.661) (0.295,0.659) (0.300,0.675) (0.305,0.673) (0.310,0.679) (0.315,0.665) (0.320,0.674) (0.325,0.680) (0.330,0.683) (0.335,0.702) (0.340,0.671) (0.345,0.650) (0.350,0.636) (0.355,0.625) (0.360,0.620) (0.365,0.610) (0.370,0.600) (0.375,0.585) (0.380,0.582) (0.385,0.581) (0.390,0.574) (0.395,0.597) (0.400,0.579) (0.405,0.582) (0.410,0.580) (0.415,0.589) (0.420,0.575) (0.425,0.580) (0.430,0.580) (0.435,0.575) (0.440,0.565) (0.445,0.559) (0.450,0.566) (0.455,0.558) (0.460,0.563) (0.465,0.560) (0.470,0.582) (0.475,0.585) (0.480,0.592) (0.485,0.583) (0.490,0.588) (0.495,0.588) (0.500,0.598) (0.505,0.594) (0.510,0.595) (0.515,0.612) (0.520,0.614) (0.525,0.638) (0.530,0.625) (0.535,0.628) (0.540,0.627) (0.545,0.635) (0.550,0.640) (0.555,0.633) (0.560,0.639) (0.565,0.634) (0.570,0.623) (0.575,0.631) (0.580,0.636) (0.585,0.649) (0.590,0.657) (0.595,0.639) (0.600,0.674) (0.605,0.676) (0.610,0.689) (0.615,0.691) (0.620,0.695) (0.625,0.691) (0.630,0.693) (0.635,0.692) (0.640,0.692) (0.645,0.681) (0.650,0.685) (0.655,0.682) (0.660,0.691) (0.665,0.715) (0.670,0.707) (0.675,0.715) (0.680,0.716) (0.685,0.714) (0.690,0.724) (0.695,0.729) (0.700,0.721) (0.705,0.714) (0.710,0.713) (0.715,0.698) (0.720,0.708) (0.725,0.720) (0.730,0.733) (0.735,0.747) (0.740,0.747) (0.745,0.749) (0.750,0.761) (0.755,0.754) (0.760,0.752) (0.765,0.765) (0.770,0.769) (0.775,0.778) (0.780,0.775) (0.785,0.781) (0.790,0.802) (0.795,0.807) (0.800,0.817) (0.805,0.828) (0.810,0.830) (0.815,0.852) (0.820,0.872) (0.825,0.868) (0.830,0.884) (0.835,0.885) (0.840,0.888) (0.845,0.903) (0.850,0.904) (0.855,0.899) (0.860,0.906) (0.865,0.936) (0.870,0.948) (0.875,0.942) (0.880,0.942) (0.885,0.955) (0.890,0.956) (0.895,0.955) (0.900,0.961) (0.905,0.969) (0.910,0.978) (0.915,0.983) (0.920,0.989) (0.925,0.992) (0.930,0.995) (0.935,0.996) (0.940,0.997) (0.945,0.999) (0.950,0.999) (0.955,1.000) (0.960,1.000) (0.965,1.000) (0.970,1.000) (0.975,1.000) (0.980,1.000) (0.985,1.000) (0.990,1.000) (0.995,1.000) (1.000,1.000) 
};
%
\addplot+[color=\tshibakacolor] coordinates { 
(0,0.25)
(0.005,0.635) (0.010,0.902) (0.015,0.897) (0.020,0.753) (0.025,0.612) (0.030,0.514) (0.035,0.515) (0.040,0.392) (0.045,0.451) (0.050,0.431) (0.055,0.429) (0.060,0.445) (0.065,0.417) (0.070,0.376) (0.075,0.412) (0.080,0.433) (0.085,0.397) (0.090,0.429) (0.095,0.446) (0.100,0.425) (0.105,0.412) (0.110,0.388) (0.115,0.360) (0.120,0.380) (0.125,0.403) (0.130,0.410) (0.135,0.428) (0.140,0.427) (0.145,0.416) (0.150,0.415) (0.155,0.416) (0.160,0.408) (0.165,0.423) (0.170,0.403) (0.175,0.377) (0.180,0.401) (0.185,0.371) (0.190,0.373) (0.195,0.377) (0.200,0.370) (0.205,0.377) (0.210,0.359) (0.215,0.360) (0.220,0.354) (0.225,0.364) (0.230,0.345) (0.235,0.372) (0.240,0.370) (0.245,0.377) (0.250,0.381) (0.255,0.389) (0.260,0.372) (0.265,0.362) (0.270,0.374) (0.275,0.372) (0.280,0.339) (0.285,0.340) (0.290,0.339) (0.295,0.341) (0.300,0.325) (0.305,0.327) (0.310,0.321) (0.315,0.335) (0.320,0.326) (0.325,0.320) (0.330,0.317) (0.335,0.298) (0.340,0.329) (0.345,0.350) (0.350,0.364) (0.355,0.375) (0.360,0.380) (0.365,0.390) (0.370,0.400) (0.375,0.415) (0.380,0.418) (0.385,0.419) (0.390,0.426) (0.395,0.403) (0.400,0.421) (0.405,0.418) (0.410,0.420) (0.415,0.411) (0.420,0.425) (0.425,0.420) (0.430,0.420) (0.435,0.425) (0.440,0.435) (0.445,0.441) (0.450,0.434) (0.455,0.442) (0.460,0.437) (0.465,0.440) (0.470,0.418) (0.475,0.415) (0.480,0.408) (0.485,0.417) (0.490,0.412) (0.495,0.412) (0.500,0.402) (0.505,0.406) (0.510,0.405) (0.515,0.388) (0.520,0.386) (0.525,0.362) (0.530,0.375) (0.535,0.372) (0.540,0.373) (0.545,0.365) (0.550,0.360) (0.555,0.367) (0.560,0.361) (0.565,0.366) (0.570,0.377) (0.575,0.369) (0.580,0.364) (0.585,0.351) (0.590,0.343) (0.595,0.361) (0.600,0.326) (0.605,0.324) (0.610,0.311) (0.615,0.309) (0.620,0.305) (0.625,0.309) (0.630,0.307) (0.635,0.308) (0.640,0.308) (0.645,0.319) (0.650,0.315) (0.655,0.318) (0.660,0.309) (0.665,0.285) (0.670,0.293) (0.675,0.285) (0.680,0.284) (0.685,0.286) (0.690,0.276) (0.695,0.271) (0.700,0.279) (0.705,0.286) (0.710,0.287) (0.715,0.302) (0.720,0.292) (0.725,0.280) (0.730,0.267) (0.735,0.253) (0.740,0.253) (0.745,0.251) (0.750,0.239) (0.755,0.246) (0.760,0.248) (0.765,0.235) (0.770,0.231) (0.775,0.222) (0.780,0.225) (0.785,0.219) (0.790,0.198) (0.795,0.193) (0.800,0.183) (0.805,0.172) (0.810,0.170) (0.815,0.148) (0.820,0.128) (0.825,0.132) (0.830,0.116) (0.835,0.115) (0.840,0.112) (0.845,0.097) (0.850,0.096) (0.855,0.101) (0.860,0.094) (0.865,0.064) (0.870,0.052) (0.875,0.058) (0.880,0.058) (0.885,0.045) (0.890,0.044) (0.895,0.045) (0.900,0.039) (0.905,0.031) (0.910,0.022) (0.915,0.017) (0.920,0.011) (0.925,0.008) (0.930,0.005) (0.935,0.004) (0.940,0.003) (0.945,0.001) (0.950,0.001) (0.955,0.000) (0.960,0.000) (0.965,0.000) (0.970,0.000) (0.975,0.000) (0.980,0.000) (0.985,0.000) (0.990,0.000) (0.995,0.000) (1.000,0.000)  
};

\addplot+[color=black] coordinates { 
(0,0.25)
(0.005,0) (1.000,0.000) 
};
\end{axis}
\end{tikzpicture}
\]
}
\caption{Alternative runs of Figure \ref{ussenatefig}, using different randomized orderings of precinct portions\label{geographicfig}}
\end{figure}

The first chart in Figure \ref{geographicfig} is similar to the one in Figure \ref{ussenatefig}. The next two are instances in which the votes that were counted  early indicated a strong showing by Tshibaka. This early lead by Tshibaka quickly evaporated as more votes were counted. Kelly and Chesbro never achieved any significant probability of winning in any of our tests.
\end{exl}

\subsection{Recount predictions}\label{recountsection}
Our algorithm can also be applied after the election results are known fully, in the case when an election seems close, and a recount is considered. The very notion of a ``close'' election can be counterintuitive in the IRV setting. Typically, the \emph{margin of victory} in any election is defined as the smallest number of votes that must be changed in order for the election outcome to become different \autocite{magrino_2011}. In a FPTP election, this margin of victory is simply the difference in vote totals between the first and second-place finishers.

It has been known for some time that determination of the margin of victory in an IRV election is not an easy problem. The natural algorithm presented in \textcite{magrino_2011} is shown to have factorial complexity in the number of candidates. Some authors have produced algorithms that run faster but give only approximate results \autocite{sarwate_2013,bhattacharyya_2021}. For 3 or 4 candidates, however, the na\"ive algorithm performs well enough.

\begin{figure}
\[
\begin{tikzpicture}
\begin{axis}[domain = -.6:.6, 
xtick={-.4,-.3, -.2, -.1,0,.1,  .2, .3, .4,.5,.6},
xticklabels={$-0.4\%$, $-0.3\%$, $-0.2\%$, $-0.1\%$, $0\%$, $0.1\%$, $0.2\%$, $0.3\%$, $0.4\%$,$0.5\%$,$0.6\%$},
tick label style={rotate=90},
xlabel=Percentage vote change per candidate after recount,
ylabel=Frequency,
height=3cm,
width=8cm,
scale only axis,
xlabel near ticks,
ylabel near ticks]
\addplot[ybar interval]  coordinates {
(-0.500,0) (-0.475,0) (-0.450,0) (-0.425,0) (-0.400,0) (-0.375,0) (-0.350,0) (-0.325,1) (-0.300,0) (-0.275,0) (-0.250,0) (-0.225,0) (-0.200,1) (-0.175,1) (-0.150,0) (-0.125,2) (-0.100,0) (-0.075,0) (-0.050,0) (-0.025,2) (0.000,9) (0.025,13) (0.050,8) (0.075,5) (0.100,6) (0.125,3) (0.150,3) (0.175,7) (0.200,3) (0.225,3) (0.250,0) (0.275,0) (0.300,0) (0.325,0) (0.350,0) (0.375,0) (0.400,2) (0.425,0) (0.450,0) (0.475,2) (0.500,0) (0.525,0) (0.550,0) (0.575,0) (0.600,0) (0.625,0) (0.650,1) (0.675,0)
};
\end{axis}
\end{tikzpicture}
\]
\caption{Histogram of historical vote shifts after recounts in US state elections, 2000--2023\label{recounthistogramfig}}
\end{figure}
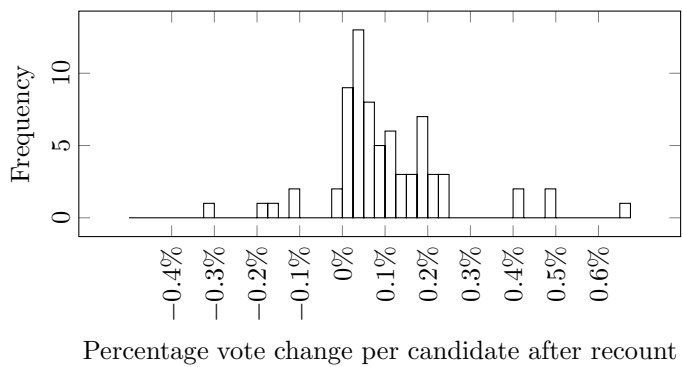

There has been some empirical study of typical recount results: we refer to the data gathered by FairVote, which collected results of all recounts in all US State elections from 2000 to 2023 \autocite{fairvote_2023}. This is comprised of 37 different elections that were recounted, almost exclusively using first-past-the-post style procedures. The data in this case shows that each recount resulted in an average change in vote totals by $0.077\%$, with a standard deviation of $0.146\%$. Plotting the various individual results in a histogram shows a distribution that appears roughly normal. See Figure \ref{recounthistogramfig}.

Thus we will take as an operating assumption that a recount will generally change the vote totals (in percentage) according to a normal distribution with mean $0.077\%$ and standard deviation $0.146\%$. 
This will allow us to estimate the probability of each candidate winning the election after the recount.

\begin{exl}\label{simplerecountex}
Consider an election between three candidates A, B, and C, in which we record a final vote tally as follows:
\[
\begin{tabular}{r|ccccccccc}
ranking: & AB & AC & BA & BC & CA & CB & A & B & C \\
\hline
votes: & 501 & 300 & 400 & 400 & 200 & 600 & 500 & 400 & 500 
\end{tabular}
\]
Calculating the IRV result, we see first-round totals of:
\[ A: 1301, \qquad B: 1200, \qquad C: 1300 \]
so B is eliminated, and the second-round totals are as follows:
\[ A: 1701, \qquad C: 1700 \]
and A is the winner. In this case, C will request a recount, and we would like to calculate the probability of each candidate winning after the recount. Intuitively, we expect A to remain the winner, with C having some nonzero chance of winning and B having no chance of winning after the recount.\footnote{In this case, we have 2401 total voters, and so our expectation is that a recount will result in a change in about $2401 \cdot 0.00077 = 1.85$ votes.} 

As described above, we assume that the vote results after the recount will be normally distributed. In this case, the probability distributions for votes after the recount are \footnote{Many intermediate rows of zeros have been omitted from this table.}:
\newcommand{\hdotline}{\hdashline[1pt/1pt]}
\[
\begin{tabular}{c|ccccccccc}
Votes & $f_{A}$ & $f_{B}$ & $f_{C}$ & $f_{AB}$ & $f_{AC}$ & $f_{BA}$ & $f_{BC}$ & $f_{CA}$ & $f_{CB}$ \\
\hline 
200 & . & . & . & . & . & . & . & 0.98 & . \\
201 & . & . & . & . & . & . & . & 0.02 & . \\
\hdotline
299 & . & . & . & . & 0.02 & . & . & . & . \\
300 & . & . & . & . & 0.79 & . & . & . & . \\
301 & . & . & . & . & 0.19 & . & . & . & . \\
\hdotline
399 & . & 0.06 & . & . & . & 0.06 & 0.06 & . & . \\
400 & . & 0.59 & . & . & . & 0.59 & 0.59 & . & . \\
401 & . & 0.34 & . & . & . & 0.34 & 0.34 & . & . \\
402 & . & 0.01 & . & . & . & 0.01 & 0.01 & . & . \\
\hdotline
499 & 0.09 & . & 0.09 & . & . & . & . & . & . \\
500 & 0.48 & . & 0.48 & 0.09 & . & . & . & . & . \\
501 & 0.38 & . & 0.38 & 0.47 & . & . & . & . & . \\
502 & 0.05 & . & 0.05 & 0.38 & . & . & . & . & . \\
503 & . & . & . & 0.05 & . & . & . & . & . \\
\hdotline
598 & . & . & . & . & . & . & . & . & 0.01 \\
599 & . & . & . & . & . & . & . & . & 0.11 \\
600 & . & . & . & . & . & . & . & . & 0.40 \\
601 & . & . & . & . & . & . & . & . & 0.38 \\
602 & . & . & . & . & . & . & . & . & 0.10 \\
603 & . & . & . & . & . & . & . & . & 0.01 \\
\end{tabular}
\]

Applying our algorithm to the distributions above gives winning probabilities as follows:
\[ A: 72.0\% \qquad B: 0\% \qquad C: 28\% \]
confirming our intuition that A and C each have a chance of winning after the recount, but B does not.
\end{exl}

We can apply the same analysis to the two races in the 2022 Alaska election that were recounted.
\begin{exl}{\bf 2022 Alaska State House District 15}

This was an election between Republicans Thomas McKay and David Eibeck, and Democrat Denny Wells, with around 7000 votes cast. We will denote these candidates respectively by M, E, and W. In the initial count for the election, Eibeck was eliminated in the first round by a wide margin, but McKay won the second round by only 7 votes. Applying the same analysis as above allows us to compute the likely results of a runoff:
\[ M: 92.9\% \qquad E: 0\% \qquad W: 7.1\% \]
and we conclude that McKay is very likely to win in the runoff, with some small chance for Wells to win.

In fact, McKay did win the runoff, winning the second round by 9 votes.
\end{exl}

\begin{exl}{\bf 2022 Alaska State Senate District E}

The 2022 race for Alaska's State Senate District E is notable because the results, as reported by the media, seemed to indicate a very close margin of loss for one candidate, who lost in the first round by only 14 votes out of about 16000 cast. The state conducted an official recount, but in fact, the election's results were not close.

This was an election between Republicans Cathy Giessel and Roger Holland, and Democrat Roselynn Cacy. We will denote these candidates respectively by G, H, and C. The typical public presentation of the results reported the following first-round totals (before the recount):
\[ G: 5652\qquad H: 5532\qquad C: 5518 \]
Here, we see Cacy behind Holland by only 14 votes, so Cacy was eliminated. Giessel went on to win the second round by a significant margin.

Despite the very slight margin between Cacy and Holland in the first round, a cursory analysis of the votes indicates that the election was not at all close. Even if Cacy had survived to the second round, she would have lost by significant margins to either Giessel or Holland, who receive far more transferred votes from one another since they are both Republicans.

Nevertheless, Cacy requested a recount. We can apply the method described above to calculate the expected results after the recount, and unsurprisingly, we obtain probabilities:
\[ G: 100\% \qquad H: 0\% \qquad C: 0\% \]
\end{exl}

\section{The IRV algorithm}
Here we review the details of how exactly the results of an IRV election are computed. We assume voters are choosing from a set of $N$ candidates $\mathcal C = \{A_1,\dots, A_N\}$. Each vote consists of an ordered list $v_1 \dots v_k$ for $0\le k\le N$ where the $v_i$ are distinct elements of $\mathcal C$. 

When a voter indicates $v_1\dots v_k$ on their ballot, this means that the candidate $v_1$ is this voter's top choice, the next candidate $v_2$ is their second choice in case $v_1$ is eliminated, and so on until $v_k$ is their last choice. If $k<N$, then this voter has left some candidates out of their vote entirely, indicating that they do not want to support those candidates under any circumstances. 

In combinatorics, an ordered subset of a given set $\cC$ is called a \emph{subset permutation} of $\cC$. We write $\Gamma(\cC)$ for the set of subset permutations of $\cC$. The number of subset permutations of a set of $N$ elements is given by the formula:
\begin{equation}\label{subsetpermcount}
\#\Gamma(\cC) = N! \sum_{k=0}^N \frac1{k!} 
\end{equation}
For various $N$, this is OEIS sequence A000522 \autocite{oeis}.
Since $\sum_{k=0}^\infty \frac1{k!}$ equals the number $e$, the size of $\Gamma(\cC)$ grows asymptotically like $eN!$ as $N$ increases.

The election results are computed recursively in rounds. In a typical round, we count the total number of times that each candidate appears as some voter's top choice. The candidate receiving the fewest of these top choice rankings is eliminated from the election. 

If there is a tie among candidates receiving the least amount, then special rules are invoked to resolve the tie which differ from place to place in real-world IRV use cases. Often in theoretical contexts, the convention is that
all candidates involved in the tie are simultaneously eliminated. In our analysis throughout the paper we consider the chance of an exact tie to be negligible, so we will assume throughout that no round ends in an exact tie.

For voters whose votes look like $v_1 \dots v_k$ with $k<N$, it is possible for all of $v_1, \dots, v_k$ to be eliminated before the end of the election. In that case, this vote is called ``exhausted'' and does not contribute to any totals in the remaining rounds.

In a real-world IRV election, the election may be ``called'' whenever one candidate is ranked as a top choice on a majority of the remaining (non-exhausted) ballots. In that case the majority candidate is mathematically guaranteed to win any future rounds, and so the ballot counting can stop before all of the rounds are computed. 

For our mathematical discussion, it will be most convenient to run the rounds to their completion in all cases. Thus, in an election with $N$ candidates (assuming no ties in any round), we will always have $N$ rounds, where the final round consists of a single candidate who is declared the winner. 

\section{Discrete convolution}
The fundamental computational tool used in our algorithm is the discrete convolution of discrete probability distributions. By \emph{discrete probability distribution} we mean a function $f:B \to \R$ where $B$ is a finite set and $f(b) \in [0,1]$ for all $b\in B$ and
\[ \sum_{b\in B}f(b) = 1. \]
For us, these values $f(b)$ will always represent probabilities associated with some random variable $X$, and so we will sometimes write $f(b)$ as $P(X=b)$.

Given two discrete probability distributions $f,g:B \to \R$, the \emph{discrete convolution} of $f$ and $g$ is defined as:
\[
f*g(k) =  \sum_{i\in B} f(i)g(k-i).
\]
If $f$ and $g$ are probability distributions of two independent random variables $X$ and $Y$, then the above becomes:
\[ 
\begin{split}
f*g(k) &= \sum_{i\in B} P(X=i) \cdot P(Y=k-i) \\
&= \sum_{i\in B} P(X=i \text{ and } Y=k-i) = P(X+Y=k). 
\end{split}
\]
This is the fundamental way in which we use the convolution: given two independent random variables, the probability distribution of their sum equals the discrete convolution.

We can also compute a convolution of more than two distributions:
\begin{equation}\label{badsum} 
f*g*h(k) = \sum_{i,j\in B} f(i)g(j)h(k-i-j) 
\end{equation}
The sum above requires a twofold iteration which may be undesirable computationally. Fortunately, the convolution satisfies an associative property:
\[ f*g*h = (f*g)*h = f*(g*h), \]
so we can find $f*g*h$ by first finding $f*g$, and then convolving $f*g$ with $h$, which is typically far more efficient than using the double sum of \eqref{badsum}. 

In fact, the summations can be avoided altogether using the property:
\[ \mathcal F (f*g) = \mathcal F (f) \cdot \mathcal F (g) \]
where $\mathcal F $ is the discrete Fourier transform, and the right side is the pointwise product. Using fast Fourier transform algorithms, this formula can be used to compute discrete convolutions more efficiently. \footnote{The na\"ive summation requires $O(n)$ operations, while the fast Fourier transform is $O(\log n)$ where $n$ is the size of $B$.}


\section{The algorithm for computing winning probabilities}\label{algorithmsection}

Our input data consists of probability distributions of random variables representing the total number of votes that will be received for each possible ranking $R\in \Gamma(\cC)$. Throughout, we make the fundamental assumption that these random variables are independent. See Section \ref{dependencesection} for a detailed discussion of the independence assumption.


Let $\mathcal E$ represent an election among a set of $N$ candidates $\mathcal C = \{A_1,\dots,A_N\}$. We assume that we are given data which consists of a set of discrete probability distributions $\{ f_{R} \mid R \in \Gamma(C) \}$, where $f_R$ is the probability distribution describing the probabilities of various vote totals for the ranking $R$.

We assume that these probability distributions all have the same discrete domain $B = b_0\mathbb N = \{0,b_0,2b_0,\dots\}$ for some positive natural number $b_0$. This set $B$ is called the \emph{set of buckets}, and the number $b_0$ is called the \emph{bucket size}. For example the example probability distributions presented in the introduction use a bucket size of 100. In that example for instance we have $f_{AC}(200) = 0.07$, which should be interpreted to mean that there is a 7\% chance that the number of votes for the ranking $AC$ is between 200 and 300. Using a smaller bucket size will allow for more specific modeling of the probability distributions, but this comes with increased computational cost since every convolution must sum over the set of buckets. The examples presented in Section \ref{alaskasection} use various different bucket sizes according to the number of votes cast in the various elections. 

Our goal is to compute the vector of winning probabilities which we denote $W(\mathcal E) \in [0,1]^N$, where coordinate $i$ of $W(\mathcal E)$ is the probability that candidate $A_i$ wins the election. As stated above, we assume that there will be no exact tie in any round.

We compute $W(\mathcal E)$ by considering all possible sequences of eliminations and their respective probabilities. For some ordered list of candidates, $\ell = v_1\dots v_k \in \Gamma(\cC)$, let $\mathcal E^\ell$ denote the election round obtained from $\mathcal E$ after eliminating the candidates $v_1, \dots, v_k$ in order (with $v_1$ eliminated first). In this context we refer to $\ell \in \Gamma(\cC)$ as an \emph{elimination order}.

We also use elimination order superscripts to unambiguously denote the random variables representing vote totals in each round. Given an elimination order $\ell\in \Gamma(\cC)$, rankings in the round $\cE^\ell$ should not use candidates appearing in $\ell$ (since those candidates have been eliminated). Thus a typical ranking in round $\cE^\ell$ is taken from the set $\Gamma(\cC \setminus \ell)$, where $\cC \setminus \ell$ is the set of all candidates from $\cC$ which do not appear in $\ell$.

For some $\ell \in \Gamma(\cC)$ and some ranking $R\in \Gamma(\cC \setminus \ell)$, let $R^\ell$ denote the random variable of the number of votes received for the ranking $R$ in the round obtained by eliminating candidates in order according to $\ell$.

For example, in an election with $\cC = \{A,B,C\}$, the number of votes for the ranking $AB$ in the first round is represented simply by the random variable $AB$, while the number of votes for the ranking $AB$ in a round after eliminating $C$ is represented by $AB^C$.

For each such random variable $R^\ell$, we write $f_R^\ell$ for its probability distribution. Thus, for example, if the probability that the ranking $AB$ receives 300 votes in the first round is $25\%$, then this would be written as $f_{AB}(300)=.25$ or $P(AB=300)=.25$.

Recall that we assume at the outset that the various random variables $R \in \Gamma(\cC)$ are independent. When $\ell \in \Gamma(\cC)$, we will want the various random variables $R^\ell \in \Gamma(\cC\setminus \ell)$ to also be independent. We give a proof of the following lemma in the appendix:
\begin{lem}\label{indeplem}
Let the random variables $\{ R \mid R \in \Gamma(\cC)\}$ be independent, and let $\ell \in \Gamma(\cC)$. Then the random variables $\{ R^\ell \mid R \in \Gamma(\cC\setminus \ell)\}$ are independent.
\end{lem}

Our computation of $W(\mathcal E)$ is done recursively by computing $W(\mathcal E^\ell)$ for all possible elimination orderings $\ell$. The recursion terminates when $\#\ell=N-1$, in which case $\ell$ includes all elements of $\mathcal C$ except for a single candidate $A_i$. This represents the final round in which only candidate $A_i$ remains, and so the vector $W(\mathcal E^\ell)$ will equal 1 in coordinate $i$, and 0 in all other coordinates.

Now we describe how to compute $W(\mathcal E^\ell)$ when $\#\ell < N-1$.
Let $E_{A_i}^\ell$ denote the probability that $A_i$ will be eliminated in election round $\mathcal E^\ell$.
Then, the desired vector of winning probabilities is computed as a weighted average:
\begin{equation}\label{recursion}
W(\mathcal E^\ell) = \sum_{A_i\in \mathcal C \setminus \ell} E_{A_i}^\ell W(\mathcal E^{\ell A_i}),
\end{equation}
where $\ell A_i$ is the elimination ordering consisting of the candidates of $\ell$ followed by $A_i$.
Since the superscript on $\mathcal E$ is longer on the right side above, eventually all superscripts will reach length $N-1$ and the algorithm will terminate. Our true goal is to compute $W(\mathcal E)$, which is done recursively as:
\begin{equation}\label{winningformula} 
W(\mathcal E) = \sum_{A_i\in \mathcal C} E_{A_i} W(\mathcal E^{A_i}). 
\end{equation}

To accomplish the computation in \eqref{recursion}, there are two ingredients: Computation of the elimination probabilities $E_{A_i}^\ell$, and computation of the probability distributions for $\mathcal E^{\ell A_i}$, given the probability distributions for $\mathcal E^\ell$. In both cases, we fundamentally use the independence of the random variables representing the vote counts. We are assuming from the outset that the vote totals in round one are independent, and the required independence will continue to hold in subsequent rounds by Lemma \ref{indeplem}.

Now we describe how to compute the various ingredients of \eqref{recursion}. 

\subsection{Computing distributions for $\cE^{\ell A}$, given those for $\cE^{\ell}$}
The easiest part is deriving the required random variables and probability distributions for $\mathcal E^{\ell A}$, given the random variables and probability distributions for $\mathcal E^\ell$. 

We need to describe how a given ranking $R^{\ell A}$ in $\mathcal E^{\ell A}$ relates to the various rankings in $\mathcal E^\ell$. Since $A$ is eliminated when we move from round $\mathcal E^\ell$ to $\mathcal E^{\ell A}$, the IRV procedure stipulates that the votes for some ranking $R^{\ell A}$ will equal the sum of all rankings from round $\mathcal E^\ell$ which become $R$ when $A$ is eliminated. Symbolically, this is written as follows:
\[ R^{\ell A} = \sum_{\substack{X \in \Gamma(\cC \setminus \ell) \\ \rho_A(X)=R}} X^\ell, \]
where $\rho_A(X)$ denotes removal of $A$ from $X$.
Because a sum of random variables corresponds to a convolution of probability distributions, we have:
\begin{equation}\label{dconvolution}
f_R^{\ell A} = \Conv_{\substack{X \in \Gamma(\cC \setminus \ell) \\ \rho_A(X)=R}} f_X^\ell. 
\end{equation}

The formula above describes how to compute the probability distributions for round $\cE^{\ell A}$, given those for round $\cE^\ell$.

\subsection{Computing elimination probabilities $E^\ell_{A_i}$}
Now we discuss the more involved computation of the elimination probabilities $E^\ell_{A_i}$ for $A_i\in \mathcal C\setminus \ell$ which appear in \eqref{recursion}. For each candidate $A_i\not\in \ell$, let $T_{A_i}^\ell$ be the random variable giving the total number of votes for which $A_i$ appears in the top position, as tallied in round $\cE^\ell$. This $T_{A_i}^\ell$ is equal to the sum of the random variables of rankings $R^\ell$ in which $R$ is a ranking beginning with $A_i$. That is,
\begin{equation}\label{Tsum}
T_{A_i}^\ell = \sum_{\substack{R\in \Gamma(\mathcal C\setminus \ell) \\ a(R)=A_i}} R^\ell, 
\end{equation}
where $a(R)\in \cC$ denotes the first-listed candidate of the ranking $R$.


Let $\tau_{A_i}^\ell$ be the probability distribution function of $T_{A_i}^\ell$. Again, since a sum of random variables corresponds to a convolution of probability distributions, we have:
\begin{equation}\label{tconvolution}
\tau_{A_i}^\ell = \Conv_{\substack{R\in \Gamma(\mathcal C\setminus \ell) \\ a(R)=A_i}} f_R^\ell.
\end{equation}

Let $\kappa^\ell_{A_i}$ be the cumulative version of $\tau^\ell_{A_i}$, that is:
\[ \kappa^\ell_{A_i} (k) = \sum_{\substack{b\in B \\ b\le k}} \tau^\ell_{A_i}(b). \]
Thus, $\kappa^\ell_{A_i} (k)$ is the probability that there are at most $k$ votes which rank $A_i$ in the top position. Accordingly, $1-\kappa^\ell_{A_i} (k)$ is the probability that there are fewer than $k$ votes which rank $A_i$ in the top position.

In a round with no ties, some candidate $A_i$ will be eliminated if there are $j$ votes placing $A_i$ in the top position while there are more than $j$ votes placing each of the other candidates in the top position. This probability is:
\[
\hat E^{\ell}_{\{A_i\}} = \sum_k P(T^\ell_{A_i}=k \text{ and } T^\ell_{A_j} > k \text{ for all }j\neq i). 
\]
The above is unwieldy as written, but our assumption of independence allows it to be computed easily. We provide proof of the following formula in the appendix.
\begin{lem}\label{elimproblem}
With the notation above, we have:
\begin{equation}\label{elimprobeq} 
\hat E^\ell_{\{A_i\}} = \sum_{k\in B} \left( \tau^\ell_{A_i}(k) \cdot \prod_{j\neq i} (1- \kappa^\ell_{A_j} (k)) \right). 
\end{equation}
\end{lem}

The quantity $\hat E^\ell_{\{A_i\}}$ is slightly less than the desired elimination probability $E^\ell_{A_i}$ appearing in \eqref{recursion}, because it represents the probability that $A_i$'s top-ranked vote bucket is less than all others. But  if the vote counts are close together, a candidate may be eliminated even if its top-ranked vote bucket equals the bucket of some other candidate. 

Recall that we consider the likelihood of actual on-the-nose ties to be negligible. But we do need to consider cases in which candidates' totals land in the same bucket. We refer to this as a ``bucket-tie''. In the context of our probabilistic model, this represents not a true tie, but a situation where the margin is ``too close to call'' in terms of our discrete probability distributions.

To handle the case where two or more candidates are bucket-tied for the least number of rankings in the top position, let $S \subseteq \{1,\dots,N\}$ be the (unordered) set indexing the bucket-tied candidates, so that $A_i$ is among those in the bucket-tie for least number of top rankings if and only if $i\in S$. Then let:
\begin{equation}\label{tieelimprobeq} 
\hat E^\ell_{S} = \sum_{k\in B} \left( \prod_{i\in S} \tau^\ell_{A_i}(k) \cdot \prod_{i\not\in S} (1- \kappa^\ell_{A_i} (k)) \right). 
\end{equation}
The above directly generalizes \eqref{elimprobeq}, and represents the probability that all candidates indexed in the set $S$ are bucket-tied for the least amount of top rankings in this round.

Since we are unable to predict the eventual resolution of a bucket-tie, we will perform each of the possible elimination outcomes and weigh them equally in the probability $E_A$. For example, if our set of candidates is $\cC = \{A_1,A_2,A_3\}$, then the total probability for elimination of $A$ in the first round is:
\[ E_{A_1} = \hat E_{\{A_1\}} + \frac12 \hat E_{\{A_1,A_2\}} + \frac12 \hat E_{\{A_1,A_3\}} + \frac13 \hat E_{\{A_1,A_2,A_3\}}, \]
where recall that $\hat E_{\{A_1\}}$ is the probability that $A_1$ alone is eliminated unambiguously, $\hat E_{\{A_1,A_2\}}$ is the probability that there is a bucket-tie between $A_1$ and $A_2$, etc. Each of the terms above can be computed by \eqref{tieelimprobeq}

In general, the proper formula for the elimination probability $E^\ell_{A_i}$ is as follows:
\begin{equation}\label{realelimprobeq} 
E^\ell_{A_i} = \sum_{\substack{S\subseteq \cC \setminus \ell}} \frac{1}{\#S} \hat E^\ell_S
\end{equation}

We have now described methods for computing each ingredient of the main formula \eqref{winningformula}, which completes the description of the algorithm.

\section{A detailed example calculation}
Before presenting an example, we give a result that can be used to simplify a given election. 

It is not hard to see that a vote in which a voter ranks all $N$ choices is equivalent to the same vote if the last choice is left blank.
\begin{lem}\label{undervotelem}
In an election with $N$ candidates, a vote $V = v_1\dots v_N$ is equivalent to the vote $V' = v_1 \dots v_{N-1}$.
\end{lem}
\begin{proof}
By ``equivalent'' we mean that substituting the vote $V'$ for $V$ will not change the first-place vote totals in any round, so that the round-by-round results of the election will be unchanged.

The votes $V$ and $V'$ are the same in each of their first $N-1$ entries. This means that the only way for $V$ and $V'$ to be tabulated differently in some rounds is if all of the candidates $v_1,\dots,v_{N-1}$ have already been eliminated. But this can only occur in round $N$, which is the final round in which the winner has already been determined, and so exchanging $V$ for $V'$ has no effect on the result.
\end{proof}

The result above means that, for a given election between $N$ candidates, we can immediately combine any votes of the form $v_1\dots v_N$ with those of the form $v_1\dots v_{N-1}$. In this way, we may assume that all voters have ranked at most $N-1$ candidates on their ballot, and none have ranked all $N$ candidates. This decreases the number of possible voter rankings by $N!$, which is a significant reduction. The number of voter rankings listing fewer than $N$ candidates is given by $N!\sum_{k=0}^{N-1} \frac1{k!}$, which is OEIS sequence A002627 \autocite{oeis}.

Lemma \ref{undervotelem} was used implicitly throughout all the examples in Section \ref{alaskasection}. For example in \eqref{FGNvotes} we have combined the actual vote totals for FNG and FN into a single column labeled FN.

\begin{exl}\label{sampleexample}
We will give the details of the full calculation for the example outlined in Section \ref{introsection}, which is a hypothetical IRV election between three candidates A, B, and C.

By Lemma \ref{undervotelem}, we may assume that no voter ranks all 3 candidates on their ballot in the first round. Thus, the possible rankings recorded on votes in this election are AB, AC, BA, BC, CA, CB, A, B, C, and $\emptyset$, where $\emptyset$ represents an exhausted vote or a vote for none of the listed candidates.

We will assume that the vote totals in the first round obey the following probability distributions:
\begin{equation}\label{simpleextable}
\begin{tabular}{c|ccccccccc}
Votes & $f_{A}$ & $f_{B}$ & $f_{C}$ & $f_{AB}$ & $f_{AC}$ & $f_{BA}$ & $f_{BC}$ & $f_{CA}$ & $f_{CB}$ \\
\hline 
0 & 0.50 & 0.10 & 0.02 & 0.01 & 0.50 & . & . & 0.09 & 0.17 \\
100 & 0.50 & 0.30 & 0.33 & 0.17 & 0.40 & . & 0.10 & 0.17 & 0.75 \\
200 & . & 0.30 & 0.21 & 0.34 & 0.07 & 0.45 & 0.30 & 0.37 & 0.08 \\
300 & . & 0.20 & 0.20 & 0.25 & 0.03 & 0.31 & 0.27 & 0.21 & . \\
400 & . & 0.10 & 0.15 & 0.13 & . & 0.20 & 0.19 & 0.10 & . \\
500 & . & . & 0.09 & 0.10 & . & 0.04 & 0.14 & 0.06 & . \\
\end{tabular}
\end{equation}
(Without Lemma \ref{undervotelem}, we would require $6$ more columns of probability distributions describing $f_{ABC}, f_{BAC},$ etc.)

First, we determine the elimination probabilities for each candidate. As a preliminary step, we use \eqref{tconvolution} to compute the probability distributions $\tau_A, \tau_B,$ and $\tau_C$, which represent probabilities for each candidate to receive a certain number of votes in the top position. By \eqref{tconvolution} we have:
\[
\begin{split}
\tau_A &= f_{AB} * f_{AC} * f_A \\
\tau_B &= f_{BA} * f_{BC} * f_B \\
\tau_C &= f_{CA} * f_{CB} * f_C
\end{split}
\]

Computing the appropriate convolutions of the columns of \eqref{simpleextable} produces the following values for these distributions, and we also compute their cumulative distributions:

\[
\begin{tabular}{r|ccc|ccc}
Votes & $\tau_A$ & $\tau_B$ & $\tau_C$ &	 $\kappa_A$ & $\kappa_B$ & $\kappa_C$ \\
\hline
0	&	0.0025	 & 	.	 & 	0.0003	& 	0.0025	 & 	0	 & 	0.0003	 \\ 
100	&	0.047	 & 	.	 & 	0.007	& 	0.0495	 & 	0	 & 	0.0073	 \\ 
200	&	0.1639	 & 	.	 & 	0.039	& 	0.2134	 & 	0	 & 	0.0463	 \\ 
300	&	0.2559	 & 	0.0045	 & 	0.095	& 	0.4693	 & 	0.0045	 & 	0.1413	 \\ 
400	&	0.2336	 & 	0.0301	 & 	0.175	& 	0.7029	 & 	0.0346	 & 	0.3163	 \\ 
500	&	0.1618	 & 	0.0867	 & 	0.1934	& 	0.8647	 & 	0.1213	 & 	0.5097	 \\ 
600	&	0.0932	 & 	0.1525	 & 	0.1813	& 	0.9579	 & 	0.2738	 & 	0.691	 \\ 
700	&	0.0337	 & 	0.1968	 & 	0.1466	& 	0.9916	 & 	0.4706	 & 	0.8376	 \\ 
800	&	0.0069	 & 	0.199	 & 	0.0931	& 	0.9985	 & 	0.6696	 & 	0.9307	 \\ 
900	&	0.0015	 & 	0.1577	 & 	0.0453	& 	1	 & 	0.8273	 & 	0.976	 \\ 
1000	&	.	 & 	0.0998	 & 	0.0181	& 	1	 & 	0.9271	 & 	0.9941	 \\ 
1100	&	.	 & 	0.0496	 & 	0.0055	& 	1	 & 	0.9767	 & 	0.9996	 \\ 
1200	&	.	 & 	0.018	 & 	0.0004	& 	1	 & 	0.9947	 & 	1	 \\ 
1300	&	.	 & 	0.0047	 & 	.	& 	1	 & 	0.9994	 & 	1	 \\ 
1400	&	.	 & 	0.0006	 & 	.	& 	1	 & 	1	 & 	1	 \\ 
\end{tabular}
\]

Then we compute the quantities of \eqref{tieelimprobeq}:
\[ 
\begin{split}
\hat E_{\{A\}} &= \sum_k \tau_A(k) (1-\kappa_B(k))(1-\kappa_C(k)) = 0.672 \\
\hat E_{\{B\}} &= \sum_k (1-\kappa_A(k)) \tau_B(k)(1-\kappa_C(k)) = 0.016 \\
\hat E_{\{C\}} &= \sum_k (1-\kappa_A(k))(1-\kappa_B(k)) \tau_C(k) = 0.167 \\
\hat E_{\{A,B\}} &= \sum_k \tau_A(k) \tau_B(k) (1-\kappa_C(k)) = 0.018 \\
\hat E_{\{A,C\}} &= \sum_k \tau_A(k) (1-\kappa_B(k)) \tau_C(k) = 0.112 \\
\hat E_{\{B,C\}} &= \sum_k (1-\kappa_A(k)) \tau_B(k) \tau_C(k) = 0.005 \\
\hat E_{\{A,B,C\}} &= \sum_k \tau_A(k) \tau_B(k) \tau_C(k) = 0.007
\end{split}
\]
and we find the elimination probabilities using \eqref{realelimprobeq}:
\begin{align*} 
\hat E_A &= \hat E_{\{A\}} + \frac12 \hat E_{\{A,B\}} + \frac12 \hat E_{\{A,C\}} + \frac13 \hat E_{\{A,B,C\}} = 0.740 \\
\hat E_B &=\hat E_{\{B\}} + \frac12 \hat E_{\{A,B\}} + \frac12 \hat E_{\{B,C\}} + \frac13 \hat E_{\{A,B,C\}} =  0.031 \\
\hat E_C &= \hat E_{\{C\}} + \frac12 \hat E_{\{A,C\}} + \frac12 \hat E_{\{B,C\}} + \frac13 \hat E_{\{A,B,C\}} =  0.229
\end{align*}

The above elimination probabilities will be used in \eqref{winningformula}. We must also recursively compute $W(\cE^A), W(\cE^B),$ and $W(\cE^C)$. We will present the details in full for $W(\cE^A)$. We want to produce a table similar to \eqref{simpleextable}, this time representing the probabilities after elimination of A. These are obtained using \eqref{dconvolution}. We have:
\[
\begin{split}
f_{BC}^A &= f_{BC} \\
f_{CB}^A &= f_{CB} \\
f_B^A &= f_{AB} * f_{BA} * f_B \\
f_C^A &= f_{AC} * f_{CA} * f_C \\
f_\emptyset^A &= f_A * f_\emptyset
\end{split}
\]

Computing these convolutions gives the desired table for the round $\mathcal E^A$:
\[
\begin{tabular}{c|ccccc}
Votes & $f^A_{\emptyset}$ & $f^A_{B}$ & $f^A_{C}$ & $f^A_{BC}$ & $f^A_{CB}$ \\
\hline 
0 & 0.50 & . & . & . & 0.17 \\
100 & 0.50 & . & 0.02 & 0.10 & 0.75 \\
200 & . & . & 0.05 & 0.30 & 0.08 \\
300 & . & 0.01 & 0.13 & 0.27 & . \\
400 & . & 0.05 & 0.18 & 0.19 & . \\
500 & . & 0.11 & 0.19 & 0.14 & . \\
600 & . & 0.18 & 0.17 & . & . \\
700 & . & 0.20 & 0.13 & . & . \\
800 & . & 0.19 & 0.08 & . & . \\
900 & . & 0.13 & 0.04 & . & . \\
1000 & . & 0.08 & 0.02 & . & . \\
1100 & . & 0.04 & . & . & . \\
1200 & . & 0.01 & . & . & . \\
\end{tabular}
\]
Note that in this round there is some expectation for a nonzero number of exhausted votes, indicated in the column labeled $f_\emptyset^A$. These votes arise from any first round votes for the ranking $A$, who has now been eliminated.

Now we can compute the distributions $\tau^A_B$ and $\tau^A_C$ using \eqref{tconvolution}:
\[
\begin{split}
\tau^A_B &= f^A_{BC} * f^A_B \\
\tau^A_C &= f^A_{CB} * f^A_C
\end{split}
\]
We perform these convolutions, construct the cumulative distributions $\kappa^A_B$ and $\kappa^A_C$, and then compute elimination probabilities using \eqref{tieelimprobeq} and \eqref{realelimprobeq}:
\[ 
\begin{split}
E^A_B &= 0.090 \\
E^A_C &= 0.909
\end{split}
\]

Now since $\mathcal E^{AB}$ and $\mathcal E^{AC}$ have only one candidate remaining, we have $W(\cE^{AB}) = (0,0,1)$ and $W(\cE^{AC}) = (0,1,0)$. Then by \eqref{recursion} we have:
\[ 
\begin{split}
W(\mathcal E^A) &= E^A_B W(\mathcal E^{AB}) + E^A_C W(\mathcal E^{AC}) \\
&= 0.090 (0,0,1) + 0.909 (0,1,0) = (0,0.909,0.090).
\end{split}
\]

Repeating the computations above for $\mathcal E^B$ and $\mathcal E^C$ in this example gives:
\[
\begin{split}
W(\mathcal E^B) &= E^B_A W(\cE^{BA}) + E^B_C W(\cE^{BC}) = (0.745, 0, 0.254) \\
W(\mathcal E^C) &= E^C_A W(\cE^{CA}) + E^C_B W(\cE^{CB}) = (0.823, 0.176, 0)
\end{split}
\]
and combining all the above in \eqref{winningformula} gives our final result of:
\[
\begin{split}
W(\cE) &= E_A W(\cE^A) + E_B W(\cE^B) + E_C W(\cE^C) \\
&= 0.740 (0,0.909,0.090) + 0.031 (0.745, 0, 0.254) + 0.229 (0.823, 0.176, 0) \\
&= (0.048, 0.861, 0.089)
\end{split}
\]
So the chances of winning for A, B, C respectively are $4.8\%$, $86.1\%$, and $8.9\%$.

Included in all of the intermediate calculations above are all the figures necessary to produce the weighted elimination tree:
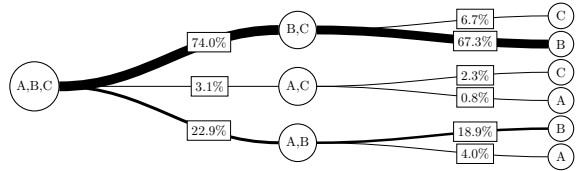
\[
\tikzset{vert/.style={draw=black,circle}}
\tikzset{weight/.style={fill=white,line width=.1,draw=black,rectangle,pos=.7}}
\begin{tikzpicture}[scale=0.5,every node/.style={scale=0.5}]
\node[vert] (ABCx) at (0,0) {A,B,C};
\node[vert] (BCxA) at (7,1.5) {B,C};
\node[vert] (CxAB) at (14,1.875) {C};
\draw[line width=0.3347833558926131] (BCxA) to[out=0,in=180] node[weight] {6.7\%} (CxAB);
\node[vert] (BxAC) at (14,1.125) {B};
\draw[line width=3.3664669099948124] (BCxA) to[out=0,in=180] node[weight] {67.3\%} (BxAC);
\draw[line width=3.701250265887426] (ABCx) to[out=0,in=180] node[weight] {74.0\%} (BCxA);
\node[vert] (ACxB) at (7,0.0) {A,C};
\node[vert] (CxBA) at (14,0.375) {C};
\draw[line width=0.11425619503727799] (ACxB) to[out=0,in=180] node[weight] {2.3\%} (CxBA);
\node[vert] (AxBC) at (14,-0.375) {A};
\draw[line width=0.03902805354479181] (ACxB) to[out=0,in=180] node[weight] {0.8\%} (AxBC);
\draw[line width=0.1532842485820698] (ABCx) to[out=0,in=180] node[weight] {3.1\%} (ACxB);
\node[vert] (ABxC) at (7,-1.5) {A,B};
\node[vert] (BxCA) at (14,-1.125) {B};
\draw[line width=0.9432595944273449] (ABxC) to[out=0,in=180] node[weight] {18.9\%} (BxCA);
\node[vert] (AxCB) at (14,-1.875) {A};
\draw[line width=0.20220589110316017] (ABxC) to[out=0,in=180] node[weight] {4.0\%} (AxCB);
\draw[line width=1.1454654855305049] (ABCx) to[out=0,in=180] node[weight] {22.9\%} (ABxC);
\end{tikzpicture}
\]

From the diagram we see that B is favored to win the election with a winning probability of 86.2\%, and further that the most likely order of eliminations is A first, followed by C.
\end{exl}

\section{Limitations and future work} 
In this section we present some limitations of the present work, and suggest some avenues for future work.

\subsection{Computational complexity}

The input data to our algorithm consists of one discrete probability distribution for each possible voter ranking.
By \eqref{subsetpermcount}, the number of these voter rankings grows asymptotically like $N!$ where $N$ is the number of candidates. Thus the complexity of our algorithm is at least factorial in $N$, and so we do not expect it to be practical when $N$ is large.

Our algorithm performs well when the number of candidates is 3 or 4. For an election with 3 candidates and domains using around 1500 vote buckets, computing the weighted elimination tree takes under 0.1 seconds on a typical consumer laptop. For 4 candidates it takes about 0.5 seconds. For 5 candidates it takes 14 seconds, and for 6 candidates, about 5 minutes. 


\subsection{The assumption of independence}\label{dependencesection}

Our algorithm fundamentally relies on the assumption that the probabilitiy distributions $f_R$ for various $R \in \Gamma(\mathcal C)$ are independent. This allows for convolutions to be used in \eqref{dconvolution} and \eqref{tconvolution}. Without the assumption of independence, these convolutions would need to be replaced by some more complicated formulas to compute joint probabilities.

The extent to which this assumption is reasonable will depend on the specific application. For the examples presented in Section \ref{alaskasection}, we believe that some dependence does in fact exist, although it is hard to judge how significant this may be. Therefore our algorithm for deriving winning probabilities should be regarded as a sort of opening salvo: we are modeling the votes in the most mathematically simple way available. Further work can improve on this basic model by introducing correlations between these distributions which are application-specific. 

For the case of predicting votes based on partial vote tallies as presented in Section \ref{unboundsec}, we suggest that correlations will exist having two main types.

\emph{Political correlations:} It is natural to expect correlations between vote tallies for candidates who are politically similar. Thus for example if candidates A and B are politically similar, and opposed to C, then we expect for the tallies for the rankings A and B to be positively correlated with each other, and each to be negatively correlated with the ranking C. Accurately modeling these correlations will require detailed analysis of the voter perception of the various candidates, presumably based on polling or other metrics.

\emph{Mathematical correlations:} We also expect purely formal correlations between various rankings which are mathematically similar. For example, in an election with candidates A,B,C,D,E, even without knowing any political information about these candidates, we can expect a correlation of the ranking ABCD with the ranking ABCE, because these two rankings are formally similar. 
We expect that these correlations could be modeled using one of the many ``string metrics'' which exist in the literature for quantifying the similarity between two strings.\footnote{For example, \citeauthor{kendall_1938}'s (\citeyear{kendall_1938}) $\tau$-distance.}

In the case of predicting recounts as in Section \ref{recountsection}, the independence assumption seems mostly justified: for instance in Example \ref{simplerecountex} the shift in votes for the ranking AB after the recount should be independent from the shift in votes for the ranking BA after the recount. There may, however, be a slight positive correlation in the absolute value of the vote shifts among all rankings when recounting ballots from the same polling place or jurisdiction. (For example if a certain polling place reports many votes changed for ranking AB after the recount, then we may take this as a sign that they are generally error prone in counting, and so we may expect higher rates of vote changes across all rankings.)

\subsection{Future Work} 
The most obvious avenue for future work will be to improve the model by allowing for correlations between the distributions. Hopefully this can be done in a way that improves the accuracy of the predictions without significantly compromising computational efficiency.

It is also natural to consider if our basic model can be adapted to the context of multi-winner contests decided by Single Transferable Vote (STV). We expect that the general idea of computing weighted averages recursively can be applied to STV as well.

If ours or similar algorithms are used for real-time prediction of in-progress elections, there is a need to present the results visually in a thoughtful and informative way. We have shown several types of visualizations of our algorithm's output. A diagram such as Figure \ref{house18fig} provides a nice way to track predictions over time as more and more votes are counted. A single application of the algorithm can produce a weighted elimination tree, which includes more information but represents a prediction of only one snapshot in time. In elections between 3 candidates a ternary chart as in Figure \ref{house18ternaryfig} may be useful. It would be nice to see more options for similar parametric diagrams when the number of candidates is more than 3.

Finally we point out that many questions still remain concerning the interpretations of the algorithm's results. For example media organizations often want to ``call the election'' when they have determined that one candidate's probability of winning is overwhelming. It is unclear in the case of our algorithm when such a ``call'' is justified. 

For example in the data presented in Figure \ref{ussenatefig}, we report a 92.7\% chance of a win by Murkowski after counting only 2.5\% of the vote. Obviously it is not appropriate to call the election at that point. In the same example, Murkowski's probability of winning reaches 100\% after 94.5\% of the votes are counted. But probably a call for Murkowski is justified earlier than 94.5\% in this case. It would be interesting if more analysis could add flares to the lines in Figure \ref{ussenatefig}, so that the election can be called with high confidence as soon as the flares are non-overlapping. 

Political analysts will need to wrestle with these and other related questions if these probabilistic methods are to be used effectively.

\section*{Appendix: Proofs of technical lemmas}
We begin with a standard fact about disjoint sums of independent random variables. The proof is elementary, but we include it for the sake of completeness.
\begin{lem}\label{sumlemma}
Let $\{X_1,\dots, X_n, Y_1,\dots, Y_m\}$ be a set of independent random variables. Then the two sums $X_1 + \dots + X_n$ and $Y_1 + \dots + Y_m$ are independent.
\end{lem}
\begin{proof}
The proof is a simple calculation. Using independence of $\{X_1,\dots, X_n, Y_1,\dots, Y_m\}$, we have:
\begin{align*}
P\left(\sum_i X_i = k \text{ and } \sum_j Y_j = \ell \right) 
&= \sum_{\substack{k_1+ \dots k_n = k \\ \ell_1 + \dots + \ell_m = \ell}} P(X_i = k_i \text{ and } Y_j = \ell_j) \\
&= \sum_{\substack{k_1+ \dots k_n = k \\ \ell_1 + \dots + \ell_m = \ell}} P(X_i = k_i) P(Y_j = \ell_j)  \\
&= \sum_{k_1 + \dots + k_n = k} P(X_i=k_i) \sum_{\ell_1+ \dots +\ell_m = \ell} P(Y_j = \ell_j) \\
&= P\left(\sum_i X_i = k \right) P\left(\sum_j Y_j = \ell \right),
\end{align*} 
and we have shown that $\sum_i X_i$ and $\sum_j Y_j$ are independent.
\end{proof}

Now we are ready to prove Lemmas \ref{indeplem} and \ref{elimproblem}.

\begin{proof}[Proof of Lemma \ref{indeplem}]
Let the random variables $\{ R \mid R \in \Gamma(\cC)\}$ be independent, and let $\ell \in \cC$, and we will show that the random variables $\{R^\ell \mid R \in \Gamma(\cC)\}$ are indepencent.

By induction on the length of $\ell$, it suffices to prove the lemma in the case when $\ell$ consists of a single candidate $\ell = A\in \cC$. Take two subset permutations $S^A, R^A \in \Gamma(\cC\setminus A)$ with $S^A \neq R^A$, and we must show that $S^A$ and $R^A$ are independent. 

As a random variable, $S^A$ is the sum of all subset-permutations $X\in \Gamma(\cC)$ for which $\rho_A(X) = S$, where $\rho_A$ denotes elimination of $A$ from $X$.
Let $\{X_1,\dots, X_s\} \in \Gamma(\cC)$ be this set of all such subset-permutations, and we have $S^A = \sum_i X_i$. 
Similarly let $Y_1,\dots,Y_t$ be the set of subset-permutations with $\rho_A(Y)=R$, and we have $R^A = \sum_i Y_i$.

Now we claim that we must have $X_i \neq Y_j$ for all $i,j$: if we had $X_i = Y_j$, then $S = \rho_A(X_i) = \rho_A(Y_j) = R$, and so $S^A = R^A$, but we have assumed that $S^A \neq R^A$. Since $X_i \neq Y_j$ for all $i,j$, and these random variables are all taken from the independent set $\{ R \mid R \in \Gamma(\cC)\}$, the various $X_i$ and $Y_j$ are all independent.

Then we may apply lemma \ref{sumlemma} to the set $\{X_1,\dots, X_s, Y_1, \dots, Y_t\}$, and we conclude that $S^A$ and $R^A$ are independent as desired.
\end{proof}

\begin{proof}[Proof of Lemma \ref{elimproblem}]
Using notations from Section \ref{algorithmsection}, we must show that 
\[ \hat E^\ell_{\{A_i\}} = \sum_k \left( \tau^\ell_{A_i}(k) \cdot \prod_{j\neq i} (1- \kappa^\ell_{A_j} (k)) \right). \]

First, we show that the random variables $T^\ell_{A_i}$ for various $A_i$ are independent. We must show that:
\[ P(T^\ell_{A_i} = m \text{ and } T^\ell_{A_j} = n) = P(T^\ell_{A_i} = m) P(T^\ell_{A_j} = n) \]
when $i\neq j$. Using \eqref{Tsum}, let $T^\ell_{A_i} = X^\ell_1 + \dots X^\ell_s$ for $X\in \Gamma(\cC)$ where $a(X_k)=A_i$ for each $k$, and similarly let $T^\ell_{A_j} = Y^\ell_1 + \dots Y^\ell_t$ for $Y\in \Gamma(\cC)$ where $a(Y_k)=A_j$ for each $k$. (Recall that $a(R)$ denotes the first-listed candidate in the ranking $R$.) Since $A_i \neq A_j$, each $X_k$ is different from each $Y_k$, and thus the set $\{X_1,\dots, X_s, Y_1,\dots, Y_t\}$ is independent by Lemma \ref{indeplem}. This means that $T_{A_i}^\ell$ and $T_{A_j}^\ell$ are independent by Lemma \ref{sumlemma}.

Now for fixed $k$, we have:
\[ 
\begin{split}
P(T^\ell_{A_i} = k &\text{ and } T^\ell_{A_j} > k \text{ for all }j\neq i) = \sum_{u>k} P(T^\ell_{A_i} = k \text{ and } T^\ell_{A_j} =u \text{ for all }j\neq i) \\
&=\sum_{u>k}  P(T^\ell_{A_i} = k) \prod_{j\neq i} P(T^\ell_{A_j} =u) = \tau^\ell_{A_i}(k) \prod_{j\neq i}\sum_{u>k} \tau^\ell_{A_j}(u) \\
&= \tau^\ell_{A_i}(k) \prod_{j\neq i} \left(1-\sum_{u\le k} \tau^\ell_{A_j}(u)\right) = \tau^\ell_{A_i}(k) \prod_{j\neq i} \left(1-\kappa^\ell_{A_j}(k) \right)
\end{split}
\]
Summing the above over $k$ gives:
\[ \begin{split}
\hat E^{\ell}_{\{A_i\}} &= \sum_k P(T^\ell_{A_i}=k \text{ and } T^\ell_{A_j} > k \text{ for all }j\neq i) \\
&= \sum_k  \tau^\ell_{A_i}(k) \prod_{j\neq i} \left(1-\kappa^\ell_{A_j}(k) \right)
\end{split}
\]
as desired.

\end{proof}

\section*{Funding}
The authors did not utilize any external funding for this project.

\section*{Acknowledgements}
The authors would like to thank Dr. Laura Dumitrescu for helpful conversations.

\section*{Data Availability Statement}
Replication code for this article is available at \url{https://github.com/cstaecker/irvprob}

\section*{Competing Interests}
The authors declare none.

\printbibliography

\end{document}